\documentclass[11pt]{article}

\usepackage{graphicx,float,wrapfig}

\usepackage{amsfonts}
\usepackage{amsmath}
\usepackage{amsthm}
\usepackage{amssymb}
\usepackage{authblk}

\usepackage[utf8]{inputenc}

\usepackage[normalem]{ulem} 
\usepackage{soul} 

\usepackage[a4paper, textwidth={17cm}]{geometry}

\newcommand{\mket}[1]{| #1 \rangle}
\newcommand{\mbra}[1]{\langle #1 |}
\newcommand{\mbraket}[2]{\langle #1 | #2 \rangle}
\newcommand{\mdenbraket}[2]{| #1 \rangle \langle #2 |}

\newcommand{\mtr}[1]{\mathrm{Tr}\left( #1 \right)}
\newcommand{\mptr}[2]{\mathrm{Tr_{#2}}\left( #1 \right)}

\newcommand{\nC}{\mathcal{C}}
\newcommand{\nH}{\mathcal{H}}

\newcommand{\nIC}{\mathcal{IC}}

\newcommand{\nHS}{\mathcal{HS}}

\newcommand{\imag}{{\bf i}}

\newtheorem*{remark}{Remark}
\newtheorem*{theorem}{Theorem}

\begin{document}

\title{Quantum qubit switch: entropy and entanglement}


\author[1]{Marek Sawerwain}
\author[2]{Joanna Wi\'sniewska}


\affil[1]{Institute of Control \& Computation Engineering University of Zielona G\'ora, Licealna 9, Zielona G\'ora 65-417, Poland; M.Sawerwain@issi.uz.zgora.pl}
\affil[2]{Institute of Information Systems, Faculty of Cybernetics, Military University of Technology, Kaliskiego 2, 00-908 Warsaw, Poland; jwisniewska@wat.edu.pl}

\date{}



\maketitle

\begin{abstract}
The quantum entanglement is considered as one of the most important notions of quantum computing. The entanglement is a feature of quantum systems and it is used as a basis for many quantum algorithms and protocols. In this paper we analyze the level of entanglement for the quantum switch, during its work. The level of entanglement during the correct operating may be compared with the situation when a noise is present in the analyzed system. The noise changes the level of quantum entanglement and we utilize this observation to evaluate if the switch works properly. We present the formulae describing the value of entanglement with use of the Concurrence measure and also the value of entropy. Both mentioned values may be utilized to detect the noise during the operating of quantum switch.  
\end{abstract}

\section{Introduction} \label{lbl:introduction:MS:JW:Entropy:2017}
 
Entropy and entanglement of quantum states \cite{Ohya1993}, \cite{Bengtsson2006} are considered as the basic notions of quantum computing \cite{Nielsen_and_Chuang}. Quantum algorithms utilize these notions in many aspects, e.g. to describe some properties of the world in the quantum level. Entropy and entanglement are also phenomena on which some known protocols are based, e.g. quantum teleportation \cite{Bennett1993, Bouwmeester1997}. Additionally, a role of quantum entanglement \cite{Horodecki2009, Otfried2009} is studied in many algorithm e.g. Grover algorithm \cite{Pan2017, Fanga2005}. In this work we calculate the time dependent levels of entropy and entanglement during the work of quantum qubit switch \cite{Ratan2007}. We present the analytical formulae describing these levels what allows to evaluate the correctness of quantum switch's functioning (these relations specify for example the level of entanglement during the faultless operating of quantum switch). Introducing a decoherence \cite{Schlosshauer2007} to the system causes changes of the entanglement's level what indicates if the switch works correctly. 
Tracing the level of entanglement in the case of the switch is especially interesting because the entanglement is not present in the analyzed system at the beginning and at the end of the process. The entanglement occurs during the operating of the switch. The presented approach based on calculating the levels of entanglement and entropy to evaluate the correctness of signal's switching in different environments can be a valuable supplement to the currently developed quantum theory of the weakest preconditions \cite{DHondt2006}, with an additional mechanism for validating the correctness of the quantum computational process.

The reminder of this paper is organized as follows. In Sect.~\ref{lbl:introduction:definition:MS:JW:Entropy:2017} basic definitions referring to quantum states and gates, which are used in other sections of this work, are presented. We also quote Schmidt theorem and allege the most important measures concerning the entanglement of quantum states. In Sect.~\ref{lbl:switch:definition:MS:JW:Entropy:2017} we discuss the notion of quantum switch and present a Hamiltonian describing the switch's dynamics. The work contains a spectral decomposition of a unitary operation of the quantum switch. There is also described a state of quantum system which characterizes the working of quantum switch in time $t$. We set the equation of Fidelity measure for the quantum switch discussed in this paper.

Sect.~\ref{lbl:sec:entanglement:level:MS:JW:Entropy:2017} presents some quantum detection techniques which allow to evaluate the level of entanglement in a moment $t$. Additional examples referring to the level of entanglement with a noise presence are presented in Sect.~\ref{lbl:qs:noise:presence:MS:JW:Entropy:2017}.

Final comments and conclusions are submitted in Sect.~\ref{lbl:conclusions:MS:JW:Entropy:2017}. The article is ended with the references section.

\section{Introductory definitions} \label{lbl:introduction:definition:MS:JW:Entropy:2017}

A basic information unit in quantum computing is a qubit. When we want to express that a~qubit has some value assigned to itself, we say about qubit's state. A state of a qubit is presented by a~normalized vector in 2-dimensional Hilbert space $\nH_2$. The definition of qubit, and its state naturally, expressed in Dirac notion as $\mket{\psi}$ with use of two normalized vector may be presented as:
\begin{equation}
\mket{\psi} = \alpha \mket{0} + \beta \mket{1},
\end{equation}
where $\alpha, \beta \in \nC$ (where $\nC$ represents the set of complex numbers) and ${\vert\alpha\vert}^2 + {\vert\beta\vert}^2 = 1$. The state $\mket{\psi}$ can be also called a vector state.

The vectors $\mket{0}, \mket{1}$ may be expressed as:
\begin{equation}
	\vert 0 \rangle =
	\left[
	\begin{array}{c}
		1\\
		0\\
	\end{array}
	\right],
	\enspace
	\vert 1 \rangle =
	\left[
	\begin{array}{c}
		0\\
		1\\
	\end{array}
	\right].
\end{equation}
These vectors are orthogonal to each other and this representation is called a standard computational basis.

As we can join classical bits together into a register and also qubits may be joined into a quantum register. In general, a state $\mket{\Psi}$ of $n$-qubit register is expressed by a tensor product:
\begin{eqnarray}
\mket{\Psi} = \mket{\psi_0} \otimes \mket{\psi_1} \otimes \mket{\psi_2} \otimes ... \otimes \mket{\psi_{n-1}} , 
\label{lbl:eqn:product:state:Entropy:2017:MS:JW}
\end{eqnarray}
where $\otimes$ stands for the operation of tensor product.

\begin{remark}
It should be stressed that some quantum states cannot be described by a tensor product like in Eq.~(\ref{lbl:eqn:product:state:Entropy:2017:MS:JW}). As an example so-called EPR state may be recalled:
\begin{equation}
\mket{\psi} = \frac{1}{\sqrt{2}}\left(\mket{00} + \mket{11}\right) .
\end{equation}
If a quantum state cannot be decomposed to a tensor product then the state is termed as entangled. The entanglement is a characteristic feature appearing in quantum systems and plays an important role in quantum computations, e.g. teleportation protocol, key distribution in quantum cryptography \cite{Ekert1991}. Detecting the presence of entanglement and also calculating the level of entanglement still constitute a very important field for research. In section (\ref{lbl:ss:detection:entanglement}) the selected measures are presented to evaluate the level of entanglement in quantum register.
\end{remark}

As a motivation to introduce density matrices we assume the case when a state of quantum system is not known completely. More precisely, only a set of states $\mket{\psi_i}$ and probabilities $p_i$ correlated with each state are known. The system is in a state described by a vector $\{ p_i, \mket{\psi_i} \}$. The set of states is an ensemble of pure states. The density operator expressing the system's state is: 
\begin{equation}
\rho = \sum_{i} p_i \mdenbraket{\psi_i}{\psi_i} .
\end{equation}
For the pure states $\mket{\psi}$ the density matrix is an operator $\mket{\psi} \mbra{\psi}$. Density matrices in this work are used in the context of introducing noise to analyzed quantum system. 

\subsection{Quantum gates}

Quantum gates represent some unitary operations. Gates acting on one qubit we call 1-qubit gates and gates affecting many qubits we call $n$-qubit gates. The quantum switch described in this work uses $n$-qubit gates but for the noise description 1-qubit Pauli gates are utilized. 

The elementary Pauli gates' set may be defined as:
\begin{eqnarray}
X = \sigma_X = \left( \begin{array}{cc} 0 & 1 \\ 1 & 0 \end{array} \right), \; Y = \sigma_Y = \left( \begin{array}{cc} 0 & i \\ -i & 0 \end{array} \right), \notag \\
Z = \sigma_Z = \left( \begin{array}{cc} 1 & 0 \\ 0 & -1 \end{array} \right), \; I = \left( \begin{array}{cc} 1 & 0 \\ 0 & 1 \end{array} \right).
\end{eqnarray}
The gate $I$ does not belong to the Pauli gates but in descriptions of $n$-qubit operations we use the symbol $I$ to represent an identity matrix. The gate $I$, placed in a quantum circuit, does not perform any operation on qubits. 

A quantum gate CNOT is a controlled negation gate (it is also called a XOR gate). This gate is needed to build a quantum switch. Formally we can describe the way the CNOT gate works as follows:
\begin{equation}
\mket{ab} = \mket{a, a \oplus b} ,
\end{equation}
where $\oplus$ stands for the addition modulo $2$ and comma is used only as a separator between first and second qubit. The CNOT gate is necessary when we want to introduce an entanglement between two qubits. A matrix form of CNOT gate is:
\begin{equation}
\mathrm{CNOT}=\left(
\begin{array}{cccc}
1 & 0 & 0 & 0 \\
0 & 1 & 0 & 0 \\
0 & 0 & 0 & 1 \\
0 & 0 & 1 & 0 \\
\end{array} \right) .
\end{equation}
If the CNOT gate affects the states from the standard computational basis, the results are:
\begin{equation}
\begin{array}{l}
\mathrm{CNOT} \mket{00} = \mket{00} \\
\mathrm{CNOT} \mket{01} = \mket{01} \\
\mathrm{CNOT} \mket{10} = \mket{11} \\
\mathrm{CNOT} \mket{11} = \mket{10}
\end{array} .
\end{equation}
The controlled negation gate may have more than one controlling line. As an example of such case we can present a Toffoli gate. A matrix representation of this gate is:
\begin{equation}
\mathrm{Toffoli}= \left(
\begin{array}{ccc}
 I &   & \\
   & I & \\
   &   &  X \\
\end{array}
\right)
=
\left(
\begin{array}{cccccccc}
1 & 0 & 0 & 0 & 0 & 0 & 0 & 0 \\
0 & 1 & 0 & 0 & 0 & 0 & 0 & 0 \\
0 & 0 & 1 & 0 & 0 & 0 & 0 & 0 \\
0 & 0 & 0 & 1 & 0 & 0 & 0 & 0 \\
0 & 0 & 0 & 0 & 1 & 0 & 0 & 0 \\
0 & 0 & 0 & 0 & 0 & 1 & 0 & 0 \\
0 & 0 & 0 & 0 & 0 & 0 & 0 & 1 \\
0 & 0 & 0 & 0 & 0 & 0 & 1 & 0 \\
\end{array}
\right) .
\end{equation}
As we can see in this case we use a block description (where I represents identity matrix and X represents negation gate) as some simplification. Formally the action that Toffoli gate performs may be described as: 
\begin{equation}
\mket{abc} = \mket{a,b,a \cdot b \oplus c} ,
\end{equation}
where $a \cdot b$ means a product of states $a$ and $b$ and again comma is used only as a separator.

For the Toffoli gate working on states from the standard basis we have: 
\begin{equation}
\begin{array}{l}
\mathrm{Toffoli} \mket{000} = \mket{000} \\
\mathrm{Toffoli} \mket{010} = \mket{010} \\
\mathrm{Toffoli} \mket{100} = \mket{100} \\
\mathrm{Toffoli} \mket{110} = \mket{111} \\
\mathrm{Toffoli} \mket{001} = \mket{001} \\
\mathrm{Toffoli} \mket{011} = \mket{011} \\
\mathrm{Toffoli} \mket{101} = \mket{101} \\
\mathrm{Toffoli} \mket{111} = \mket{110}
\end{array} .
\end{equation}
The negation of the third state is carried out only when the two first qubits are in the state $\mket{1}$.

\begin{figure}
\begin{center}
\includegraphics[height=3.75cm]{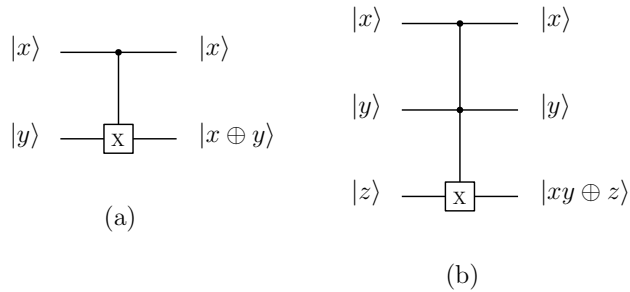}
\end{center}
\caption{The graphical representations of gates: CNOT (a) and Toffoli (b)}
\label{lbl:fig:CNOT:Toffoli:MS:JW:Entropy:2017}
\end{figure}

Fig.~\ref{lbl:fig:CNOT:Toffoli:MS:JW:Entropy:2017} depicts the graphical representations of gates: CNOT and Toffoli. Naturally, in literature we can meet also other graphical representations of these gates, but in this work we use the symbols as at mentioned Fig.~\ref{lbl:fig:CNOT:Toffoli:MS:JW:Entropy:2017}.

\subsection{Detection of quantum entanglement} \label{lbl:ss:detection:entanglement}

If we consider analyzing the entanglement of quantum states, we ought to mention the Schmidt decomposition theorem.

For a finite-dimensional Hilbert space $\nH$ correlated Hilbert-Schmidt space is denoted as $\nHS(\nH)$. The space $\nHS(\nH)$ consists of linear operations, carried out in space $\nH$, correlated with a scalar product: 
\begin{equation}
\mbraket{A}{B}_{ \nHS } = \mtr{ A^{\dagger} B } .
\end{equation}
A set of $(E_i)$, $i=1,\ldots,\dim(\nH)^2$, linearly independent operators in a Hilbert space $\nH $ is called a complete orthonormal set if and only if $\mbraket{E_i}{E_j}_{HS}=\delta_{i,j}$, where $j=1,\ldots,\dim(\nH)^2$. If also all operators $E_i$ are Hermitian then the set $(E_i)$ is called a complete Hermitian orthonormal set.

For quantum states the following theorem may be presented:

\begin{theorem} (Schmidt Decomposition) \\
Let $\rho \in (\mathcal{H}_A \otimes \mathcal{H}_B)$ then exists a number $r > 0$ (called a Schmidt rank of a state $\rho$) and complete orthonormal set $(E_i^A)$ (and also $(E_j^B)$) in space $\nHS(\mathcal{H}_A)$ (respectively $\nHS(\mathcal{H}_B)$) such that 
\begin{equation}
\rho = \sum_{\alpha=1}^r \lambda_{\alpha} E^A_{\alpha} \otimes E^B_{\alpha} ,
\end{equation}
where scalars $\lambda_{\alpha} > 0$ are termed as Schmidt coefficients of a state $\rho$. If all numbers $\lambda_{\alpha}$ are different, then the decomposition is uniquely determined.
\label{lbl:eqn:schmidt:decomposition:MS:JW:Entropy:2017}
\end{theorem}

\begin{remark}
In \cite{Terhal2000, Terhal2002} various definitions for density matrices are discussed. The Schmidt decomposition presented above, Schmidt rank $r$, Schmidt coefficients $\lambda_{\alpha}$ and an orthonormal set directly correspond to the state $\rho$. It means that the properties of $\rho$ i.e. whether the state is separable or entangled may be gained from the mentioned decomposition. Especially for a pure state $\rho$, if the Schmidt rank r=1 then $\rho$ is a separable state. For $r>1$ the state $\rho$ is entangled and if all $\lambda_{\alpha}$ are non-zero and equal to one another then $\rho$ is maximally entangled state.
\end{remark}

In \cite{Peres1996} the Peres-Horodecki separability criterion (also termed as Positive Partial Transpose (PPT) criterion) was presented which refers to the presence of entanglement in bipartite systems sized $2 \times 2$ i.a. for system of two qubits (the PPT criterion is also true for systems sized $2 \times 3$ and $3 \times 2$). The bipartite system $\rho_{AB}$ is separable when the following relation is true:
\begin{equation}
\rho^{T_A}_{AB} \equiv (T \otimes I) \rho_{AB} = \sum_{j} p_j (\rho_j)^{T_A}_{A} \otimes (\rho_j)_B \geq 0,
\end{equation}
where $T^A$ is a partial transposition operation carried out on the first qubit $A$, i.a. $(T \otimes I)$. If $\rho^{T_A}_{AB} \geq 0$ the eigenvalues of the state $\rho^{T_A}_{AB}$ are positive, so the state $\rho^{T_A}_{AB}$ is separable. If there is a negative eigenvalue, the examined state is entangled.

To evaluate the level of entanglement, because we analyze a two-qubit system, the Concurrence measure is used:
\begin{equation}
\nC( \rho )  = \max \left( 0, \lambda_1 - \lambda_2 - \lambda_3 - \lambda_4 \right) ,
\label{lbl:eq:concurrency:MS:JW:EntropyArt2017}
\end{equation}
where $\lambda_i$ and $i \in \{1,2,3,4 \}$, are eigenvalues of the Hermitian matrix $R_\rho$ (given in descending order):
\begin{equation}
R_\rho  = \sqrt{ \sqrt{\rho} \tilde{\rho}  \sqrt{\rho}  }.
\end{equation}
The state $\tilde{\rho}$ we obtain as:
\begin{equation}
\tilde{\rho} = (\sigma_y \otimes \sigma_y) \rho^{\star} (\sigma_y \otimes \sigma_y),
\end{equation}
where $\rho^{\star}$ denotes the complex conjugate. More details concerning above criterion may be found in \cite{Wootters1998}.

The Concurrence (also called as Generalized Concurrence or I-Concurrence)  measure may be also described as:
\begin{equation}
\nIC(\mket{AB})  = \sqrt{2 \left( 1 - \mtr{\sigma^2_A} \right)} ,
\label{lbl:eq:general:concurrency:MS:JW:EntropyArt2017}
\end{equation}
where $\sigma^2_A$ is a square of the state matrix after the operation of partial trace where the $\mket{B}$ state was traced out.

\subsection{Entropy for entanglement detection}

The value of entropy may be utilized to evaluate the level of entanglement of a quantum state. Generally, for the density matrix $\rho$ the von Neumann entropy $S$ may be described as:
\begin{equation}
S( \rho ) = - \mtr{ \rho \ln \rho } .
\end{equation}
Entropy may be useful for detection of entanglement in a bipartite quantum system. For a pure state $\mket{\psi}$:
\begin{equation}
\rho_{AB} = {\mdenbraket{\psi}{\psi}}_{AB} ,
\end{equation}
the entropy may be calculated e.g. for the part $A$:
\begin{equation}
S( \rho_{A} ) = -\mtr{\rho_A \log \rho_A} = -\mtr{\rho_B \log \rho_B} = S( \rho_{B} ) ,
\label{lbl:eqn:entropy:bipartite:identitity}
\end{equation}
and also for the part $B$ -- these values are equal. Naturally, $\rho_{A} = \mptr{\rho_{AB}}{B}$ and  $\rho_{B} = \mptr{\rho_{AB}}{A}$ are the partial trace reduced density matrices. Because of (\ref{lbl:eqn:entropy:bipartite:identitity}) the value of entropy, in the context of entanglement, shows the amount of entangled information between parts $A$ and $B$.

\begin{remark}
If for a matrix $\rho_A$ we will perform a spectral decomposition:
\begin{equation}
\rho_A = \sum_{k} \lambda_k \mket{k}\mbra{k},
\end{equation}
where $\lambda_k$ are eigenvalues and $\mket{k}$ are eigenvectors of matrix $\rho_A$, then the entropy may be calculated as:
\begin{equation}
S(\rho_A) = - \sum_{k} \lambda_k \log \lambda_k = \sum_k \lambda_k \log \frac{1}{\lambda_k} .
\end{equation}
\end{remark}

\subsection{Decoherence modeled as quantum channels}

In Sect.~\ref{lbl:qs:noise:presence:MS:JW:Entropy:2017} we present an influence of modeled noise on the level of entanglement. The noise may be generated by four quantum channels: Phase Flip, Bit Flip, Amplitude Damping and Phase Damping (the detailed definitions of these channels as Kraus operators are contained in Tab.~\ref{lbl:tbl:quantum:channels:MS:JW:Entropy:2017}).

\begin{table}
\begin{displaymath}
\begin{array}{|l|c|c|}
\hline
\multicolumn{1}{|c|}{\mathrm{Name \; of \; channel}} & \multicolumn{2}{c|}{\mathrm{Kraus \; operators}} \\  \hline
\mathrm{Phase \; Flip (PF)}  & E_0 = \sqrt{p} I & E_1 = \sqrt{1-p} \sigma_z \\ \hline
\mathrm{Bit \; Flip (BF)}  & E_0 = \sqrt{p} I & E_1 =  \sqrt{1-p} \sigma_x \\ \hline
\mathrm{Amplitude \; Damping (AD)}  & E_0 = \left( \begin{array}{cc} 1 & 0 \\  0 & \sqrt{1-p}  \end{array} \right) & E_1 =  \left( \begin{array}{cc} 0 & \sqrt{p} \\  0 & 0  \end{array} \right) \\ \hline
\mathrm{Phase \; Damping (PD)}  & E_0 = \left( \begin{array}{cc} 1 & 0 \\  0 & \sqrt{1-p}  \end{array} \right) & E_1 =  \left( \begin{array}{cc} 0 & 0 \\  0 & \sqrt{p}  \end{array} \right) \\ \hline
\end{array}
\end{displaymath}
\caption{Kraus operators for four quantum channels, the $p$ parameter describes the decoherence probability
}
\label{lbl:tbl:quantum:channels:MS:JW:Entropy:2017}
\end{table}
The Kraus operators were used in this work because they allow to describe the influence of the environment on an examined quantum state. If the state is presented as a density matrix $\rho$ then: 
\begin{equation}
\mathcal{K}(\rho) = \sum_{k=0}^{K-1} M_k \rho M^{\dagger}_{k},
\end{equation}
where $M_k$ are Kraus operators. The number of Kraus operators is $K < N^2$ where $N$ stands for the system's dimension. The operators $M_k$ meet the condition $\sum_{k=0}^{K-1} M^{\dagger}_k M_k = I$ and also preserve: linearity, trace and hermicity. The operators $M_k$ are completely positive, too. Further information concerning Kraus operators may be found in \cite{Kraus1983}, \cite{Verstraete2002}.

\section{Quantum switch for qubits} \label{lbl:switch:definition:MS:JW:Entropy:2017}

The work \cite{Ratan2007} presents the notion of a quantum switch as a 3-qubit controlled swap gate. Let us describe a state of this quantum system as:
\begin{equation}
\mket{\Psi_{qs}} = \mket{A}\mket{B}\mket{C} ,
\end{equation}
where the first qubit $\mket{A}$ and the second one $\mket{B}$ take the unknown initial values. The value assigned to the third controlling qubit $\mket{C}$ is $\mket{0}$ or $\mket{1}$.

We assume that the states $\mket{A}$ and $\mket{B}$ may be presented as:
\begin{equation}
\mket{A} = \alpha_{0} \mket{0} + \beta_{0} \mket{1},  \;\;\; \mket{B} = \alpha_{1} \mket{0} + \beta_{1} \mket{1} ,
\end{equation}
where $\alpha_0, \beta_0, \alpha_1, \beta_1 \in \nC$ and ${\vert\alpha_0\vert}^2 + {\vert\beta_0\vert}^2 = 1$, ${\vert\alpha_1\vert}^2 + {\vert\beta_1\vert}^2 = 1$.

Citing \cite{Ratan2007} the quantum switch is some kind of controlled SWAP gate. The mentioned gate swaps the states $\mket{A}$ and $\mket{B}$ according to a value of $\mket{C}$.

The way of operating for the quantum switch may be described by two cases. The first case takes place when the state of the qubit $\mket{C}$ is $\mket{0}$:
\begin{equation}
\mket{A}\mket{B}\mket{0} \Rightarrow \mket{A}\mket{B}\mket{0} .
\end{equation}
The quantum switch does not swap the states of input qubits. The operation of swapping is connected with the second case when the state of the qubit $\mket{C}$ is $\mket{1}$:
\begin{equation}
\mket{A}\mket{B}\mket{1} \Rightarrow \mket{B}\mket{A}\mket{1},
\label{lbl:eq:swap}
\end{equation}
as it can be seen in Eq.(\ref{lbl:eq:swap}), the states of qubits $\mket{A}$ and $\mket{B}$ were swapped.

\begin{figure}
\begin{center}
\includegraphics[height=3.75cm]{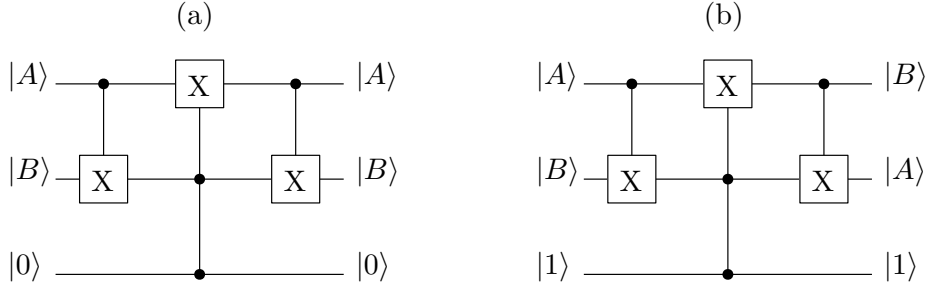}
\end{center}
\caption{The circuits illustrating the operation of quantum switch for qubits. If the state of controlling qubit is $\mket{0}$ (case (a)) the switch does not change the order of first two input states. When the state of the third qubit is expressed as $\mket{1}$ (case (b)) the quantum switch swaps the input states $\mket{A}$ and $\mket{B}$}
\label{lbl:fig:QS:EE:MS:JW:EtropyArt2017}
\end{figure}

A unitary operation corresponding to such behaviour needs using only three quantum gates i.e. two controlled negation gates and Toffoli gate. Fig.~(\ref{lbl:fig:QS:EE:MS:JW:EtropyArt2017}) depicts the quantum circuit realizing operations performed by the quantum switch.

Naturally, if the circuit and matrix forms for utilized operations are given then the matrix form of unitary operation $U_{qs}$ realizing the quantum switch may be calculated:
\begin{equation}
U_{qs} = \left(
\begin{array}{cccccccc}
 1 & 0 & 0 & 0 & 0 & 0 & 0 & 0 \\
 0 & 1 & 0 & 0 & 0 & 0 & 0 & 0 \\
 0 & 0 & 1 & 0 & 0 & 0 & 0 & 0 \\
 0 & 0 & 0 & 0 & 0 & 1 & 0 & 0 \\
 0 & 0 & 0 & 0 & 1 & 0 & 0 & 0 \\
 0 & 0 & 0 & 1 & 0 & 0 & 0 & 0 \\
 0 & 0 & 0 & 0 & 0 & 0 & 1 & 0 \\
 0 & 0 & 0 & 0 & 0 & 0 & 0 & 1 \\
\end{array}
\right) .
\end{equation}

Although the operation $U_{qs}$ captures the complete working of the switch, the evaluation of entanglement level and changes of entropy values needs a Hamiltonian presenting the switch. The switch realizes swap operation only if the third qubit is in the state $\mket{1}$. That leads to the direct form of the Hamiltonian, describing the dynamics of the operation performed by the switch:

\begin{equation}
H_{qs} = \mket{011}\mbra{101} + \mket{101}\mbra{011}  = \left(
\begin{array}{cccccccc}
 0 & 0 & 0 & 0 & 0 & 0 & 0 & 0 \\ 
 0 & 0 & 0 & 0 & 0 & 0 & 0 & 0 \\ 
 0 & 0 & 0 & 0 & 0 & 0 & 0 & 0 \\ 
 0 & 0 & 0 & 0 & 0 & 1 & 0 & 0 \\ 
 0 & 0 & 0 & 0 & 0 & 0 & 0 & 0 \\ 
 0 & 0 & 0 & 1 & 0 & 0 & 0 & 0 \\ 
 0 & 0 & 0 & 0 & 0 & 0 & 0 & 0 \\ 
 0 & 0 & 0 & 0 & 0 & 0 & 0 & 0 \\ 
\end{array}
\right) .
\end{equation}

Because the Hamiltonian $H_{qs}$ is a hermitian operator, introducing a time variable $t$ allows to express the dynamics of switch as a unitary time evolution operator:
\begin{equation}
U_{qs}(t) = e^{- \imag t H_{qs}}.
\end{equation}

A matrix form of the operator, for real values of $t$ variable, is:
\begin{equation}
U_{qs}(t) = \left(
\begin{array}{cccccccc}
 1 & 0 & 0 & 0 & 0 & 0 & 0 & 0 \\
 0 & 1 & 0 & 0 & 0 & 0 & 0 & 0 \\
 0 & 0 & 1 & 0 & 0 & 0 & 0 & 0 \\
 0 & 0 & 0 & \cos (t) & 0 & -\imag \sin (t) & 0 & 0 \\
 0 & 0 & 0 & 0 & 1 & 0 & 0 & 0 \\
 0 & 0 & 0 & -\imag \sin (t) & 0 & \cos (t) & 0 & 0 \\
 0 & 0 & 0 & 0 & 0 & 0 & 1 & 0 \\
 0 & 0 & 0 & 0 & 0 & 0 & 0 & 1 \\
\end{array}
\right),
\label{lbl:eqn:u:qs:oper}
\end{equation}
where $t \in \langle 0, \frac{\pi}{2} \rangle$. For $t= \frac{\pi}{2}$ the switch ends correctly the swap operation for input states.

We can calculate a spectral decomposition of the unitary operation given in (\ref{lbl:eqn:u:qs:oper}). In this case we obtain the following sequence of eigenvalues:
\begin{equation}
\lambda_{k} = [ -1, 0, 0, 0, 0, 0, 0, 1 ],
\label{lbl:eqn:lambdas:sequence:MS:JW:Entropy:2017}
\end{equation}
where $k \in \{0, 1, \dots, 7\}$.

That allows to describe the operation $U_{qs}(t)$ as:
\begin{gather}
U_{qs}(t) = \frac{4 e^{(\imag t \lambda_0)} }{8} \left(
\begin{array}{cccccccc}
 0 & 0 & 0 & 0 & 0 & 0 & 0 & 0 \\
 0 & 0 & 0 & 0 & 0 & 0 & 0 & 0 \\
 0 & 0 & 0 & 0 & 0 & 0 & 0 & 0 \\
 0 & 0 & 0 & 1 & 0 & 1 & 0 & 0 \\
 0 & 0 & 0 & 0 & 0 & 0 & 0 & 0 \\
 0 & 0 & 0 & 1 & 0 & 1 & 0 & 0 \\
 0 & 0 & 0 & 0 & 0 & 0 & 0 & 0 \\
 0 & 0 & 0 & 0 & 0 & 0 & 0 & 0 \\
\end{array}
\right) + \notag \\
\sum_{\stackrel{k=0}{{k \neq 4};{k \neq 6}}}^{7} \frac{4 e^{(\imag t \lambda_k)} }{8} \mdenbraket{k}{k} +
\frac{4 e^{(\imag t \lambda_7 )} }{8} \left( \begin{array}{cccccccc}
 0 & 0 & 0 & 0 & 0 & 0 & 0 & 0 \\
 0 & 0 & 0 & 0 & 0 & 0 & 0 & 0 \\
 0 & 0 & 0 & 0 & 0 & 0 & 0 & 0 \\
 0 & 0 & 0 & 1 & 0 & -1 & 0 & 0 \\
 0 & 0 & 0 & 0 & 0 & 0 & 0 & 0 \\
 0 & 0 & 0 & -1 & 0 & 1 & 0 & 0 \\
 0 & 0 & 0 & 0 & 0 & 0 & 0 & 0 \\
 0 & 0 & 0 & 0 & 0 & 0 & 0 & 0 \\
\end{array}
\right) .
\end{gather}
The values $\lambda_i$ appear in order given by the sequence (\ref{lbl:eqn:lambdas:sequence:MS:JW:Entropy:2017}) and $t \in \langle 0, \frac{\pi}{2} \rangle$.

If the unitary operation (\ref{lbl:eqn:u:qs:oper}) is used then the system's state (with the control qubit in the state $\mket{0}$) may be expressed as:
\begin{equation}
U_{qs}(t) \mket{\Psi_{qs}} = \left(
\begin{array}{c}
 \alpha _0 \alpha _1 \\
 0 \\
 \alpha _0 \beta _1 \\
 0 \\
 \alpha _1 \beta _0 \\
 0 \\
 \beta _0 \beta _1 \\
 0 \\
\end{array}
\right) .
\end{equation}
As we may notice, there is no swapping of states. The gate $U_{qs}(t)$ does not perform any action on the quantum state. The action is performed when the state of control qubit is $\mket{1}$:
\begin{equation}
U_{qs}(t) \mket{\Psi_{qs}} = \mket{\Psi^{U_{qs}(t)}_{qs}} = \left(
\begin{array}{c}
 0 \\
 \alpha _0 \alpha _1 \\
 0 \\
 \cos (t) \alpha _0 \beta _1 - \imag \sin (t) \alpha _1 \beta _0 \\
 0 \\
 \cos (t) \alpha _1 \beta _0 - \imag \sin (t) \alpha _0 \beta _1 \\
 0 \\
 \beta _0 \beta _1 \\
\end{array}
\right) ,
\label{lbl:eqn:QS:qstate:MS:JW:Entropy:2017}
\end{equation}
where $t \in \langle 0, \frac{\pi}{2} \rangle$.

The value of Fidelity measure may be calculated for the quantum switch. If we act on pure states then the Fidelity measure is expressed as:
\begin{equation}
F(\phi, \psi) = | \mbraket{\phi}{\psi} | .
\end{equation}
If the control qubit of the switch is in the state $\mket{1}$ and the states $\mket{A}$, $\mket{B}$ are unknown then the value of Fidelity measure is:
\begin{equation}
F_{qs}(t) = \left| 
  \alpha _0^2 \left(\alpha _1^2 - \imag \sin (t) \beta _1^2\right) + 2 \cos (t) \alpha _0 \alpha _1 \beta _0 \beta _1  + \beta _0^2 \left( -\imag \sin (t) \alpha _1^2 + \beta _1^2 \right) \right| .
\end{equation}
The basic algebraic transformations for $t=\frac{\pi}{2}$ show that in the end we calculate a norm of a complex number. Because of the normalization condition this value is equal to 1. The mentioned conclusion may be presented in a simpler way if the value of Fidelity measure is calculated for $\mket{A01}$ or $\mket{A11}$:
\begin{equation}
F^{A01}_{qs}(t) = \alpha _0^2 + \sin (t) \beta _0^2, \;\;\; F^{A11}_{qs}(t) =  \sin (t) \alpha _0^2 + \beta _0^2 .
\end{equation}
In these cases the values of Fidelity measure depend on the value of sine function which directly depends on $t$.

\begin{figure}
\begin{center}
\includegraphics[height=5.75cm]{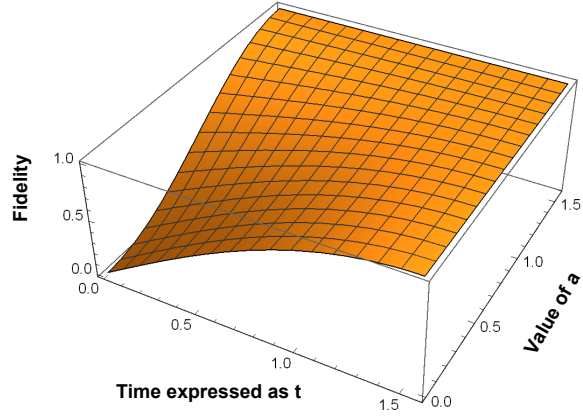}
\end{center}
\caption{The change of Fidelity value for the quantum switch in state $\mket{A01}$,  time interval $t \in \langle 0, \frac{\pi}{2} \rangle$}
\label{lbl:fig:FidelityA01:changing:EE:MS:JW:EtropyArt2017}
\end{figure}

The change of Fidelity value for state $\mket{A}$:
\begin{equation}
\mket{A} = \sin(a) \mket{0} + \cos(a) \mket{1},
\label{eqn:state:A:MS:JW:Entropy:2017}
\end{equation}
for $a \in \langle 0, \frac{\pi}{2} \rangle$ is shown at Fig.~(\ref{lbl:fig:FidelityA01:changing:EE:MS:JW:EtropyArt2017}). According to our expectations, in the moment $t=\frac{\pi}{2}$, when the switch finishes the process of information passing, the value of Fidelity measure is equal to one. 

\section{Detection and the level of entanglement} \label{lbl:sec:entanglement:level:MS:JW:Entropy:2017}

During the operation of quantum switch the phenomenon of entanglement should appear in the system. As the first step of entanglement's analysis it should be stated if the entanglement is present. In this work we act on pure states, so utilizing the Thm.~(\ref{lbl:eqn:schmidt:decomposition:MS:JW:Entropy:2017}) clearly indicates if analyzed state is entangled.

Naturally, we are interested in the case when the switch changes the information so the third qubit is in the state $\mket{1}$. The Schmidt decomposition, according to Thm.~(\ref{lbl:eqn:schmidt:decomposition:MS:JW:Entropy:2017}), is performed on a vector state which may be described as:
\begin{gather}
\left(
\begin{array}{c}
 \alpha _0 \\
 -\imag \sin (t) \beta _0 \\
 \cos (t) \beta _0 \\
 0 \\
\end{array}
\right).
\label{eqn:vector:state:for:Schmidt:decomp:MS:JW:Entropy:2017}
\end{gather}
The state presented in Eq.~(\ref{eqn:vector:state:for:Schmidt:decomp:MS:JW:Entropy:2017}) is derived from the state $\mket{A01}$ where $A$ is an unknown state of qubit, i.a. $\mket{A} = \alpha_0 \mket{0} + \beta_0 \mket{1}$. The third qubit was removed from the system, but initially its value was $\mket{1}$.

\begin{remark}
If the partial trace operation on the last qubit was performed on the state $\mket{A01}$, the density matrix of obtained state (\ref{eqn:vector:state:for:Schmidt:decomp:MS:JW:Entropy:2017}) is:
\begin{gather}
\mptr{A01}{C} = \mket{A0} = \left(
\begin{array}{cccc}
 \left(\alpha _0\right){}^* \alpha _0 & \imag \left(\sin (t) \beta _0\right){}^* \alpha _0 & \left(\cos (t) \beta _0\right){}^* \alpha _0 & 0 \\
 -\imag \left(\alpha _0\right){}^* \sin (t) \beta _0 & \left(\sin (t) \beta _0\right){}^* \sin (t) \beta _0 & -\imag \left(\cos (t) \beta _0\right){}^* \sin (t) \beta _0 & 0 \\
 \left(\alpha _0\right){}^* \cos (t) \beta _0 & \imag \left(\sin (t) \beta _0\right){}^* \cos (t) \beta _0 & \left(\cos (t) \beta _0\right){}^* \cos (t) \beta _0 & 0 \\
 0 & 0 & 0 & 0 \\
\end{array}
\right) = \rho_{A0}.
\label{eqn:state:for:Schmidt:decomp:MS:JW:Entropy:2017}
\end{gather}
It is easy to verify by the direct calculations that the density matrix~(\ref{eqn:state:for:Schmidt:decomp:MS:JW:Entropy:2017}) may be derived from the state~(\ref{eqn:vector:state:for:Schmidt:decomp:MS:JW:Entropy:2017}).
\end{remark}

Performing the Schmidt decomposition on the state~(\ref{eqn:vector:state:for:Schmidt:decomp:MS:JW:Entropy:2017}) allows to obtain the following values of coefficients $\lambda_{1}$, $\lambda_{2}$:
\begin{gather}
\lambda_0 = \frac{\sqrt{1 - \sqrt{1-{\left| \beta_0 \right|}^4 \sin^2(2 t)}}}{\sqrt{2}}, \;\;\; \lambda_1 = \frac{\sqrt{1 + \sqrt{1-{\left| \beta_0 \right|}^4 \sin^2(2 t)}}}{\sqrt{2}} .
\end{gather}

\begin{figure}
\begin{center}
\includegraphics[height=3.75cm]{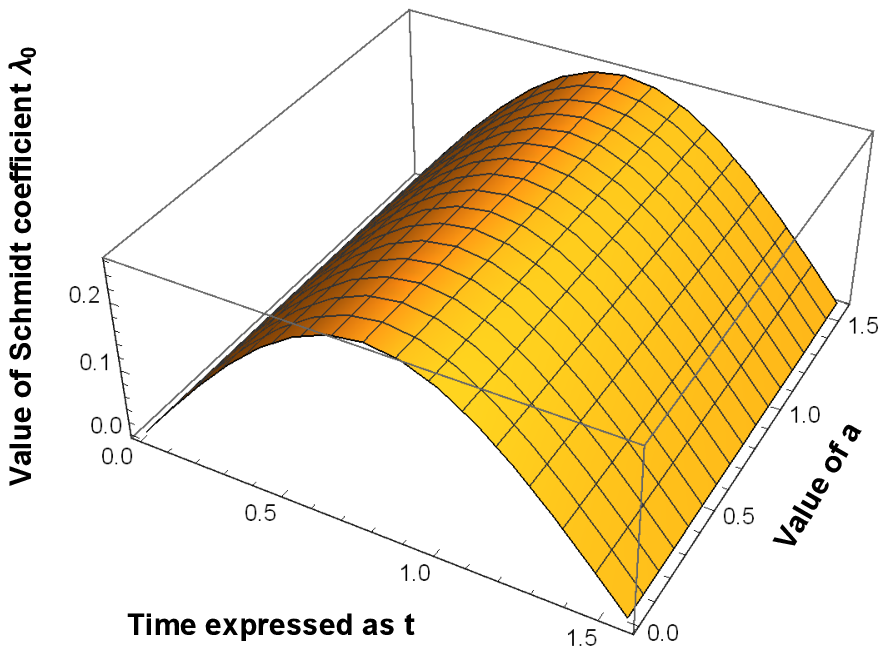} \ \includegraphics[height=3.75cm]{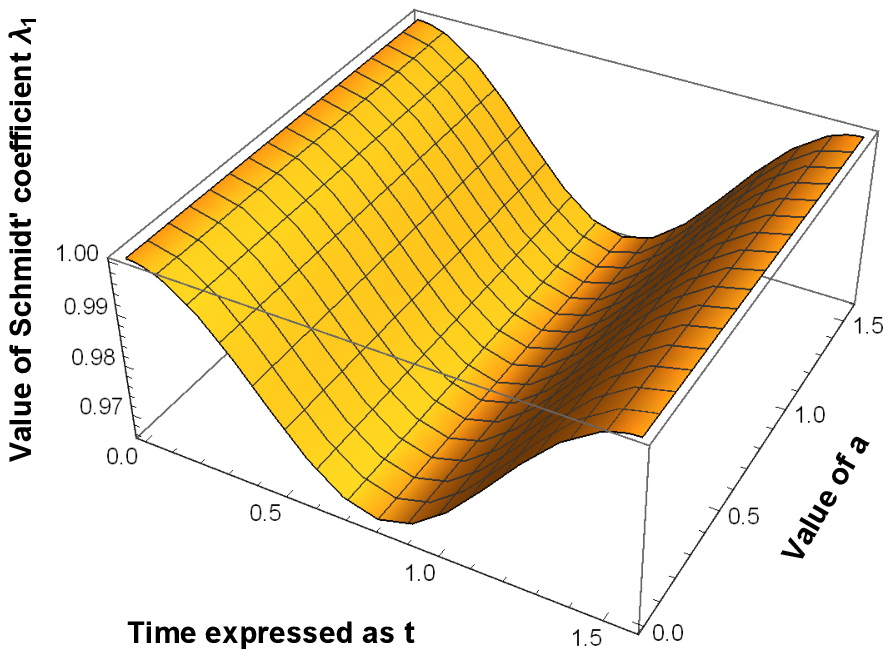}
\end{center}
\caption{The changes in Schmidt coefficients' values $\lambda_0$, $\lambda_1$ for the state~(\ref{eqn:vector:state:for:Schmidt:decomp:MS:JW:Entropy:2017}) where the state of the first qubit is described by Eq.~(\ref{eqn:state:A:MS:JW:Entropy:2017}). The values of the parameter $a$ and time $t$ are real numbers in range $\langle 0, \frac{\pi}{2} \rangle$
}
\label{lbl:fig:L0L1:changing:EE:MS:JW:EtropyArt2017}
\end{figure}

According to the expectations{\color{green},} the entanglement vanishes for $t=0$ and $t=\frac{\pi}{2}$, because for these values of $t$ the value of $\lambda_0$ is equal to zero. Theorem~(\ref{lbl:eqn:schmidt:decomposition:MS:JW:Entropy:2017}) informs that the presence of entanglement is connected with the situation when at least two Schmidt coefficients are non-zero. The changes of values of the coefficients $\lambda_0$ and $\lambda_1$ may be visualized assuming that the analyzed state is $\mket{A01}$ and the state of the first qubit $\mket{A}$ is as in Eq.~(\ref{eqn:state:A:MS:JW:Entropy:2017}).
Fig.~(\ref{lbl:fig:L0L1:changing:EE:MS:JW:EtropyArt2017}) presents the visualization of values' changes for the Schmidt coefficients for the state (\ref{eqn:state:for:Schmidt:decomp:MS:JW:Entropy:2017}).

The Schmidt decomposition is performed on a vector state. However, after adding some noise, modeled as a quantum channel, the state of the system will be described by a density matrix. Because we operate on bipartite system, the PPT criterion allows to determine the presence of entanglement. 

Calculating partial transposition on the second qubit of state $\rho_{A0}$, given by (\ref{eqn:state:for:Schmidt:decomp:MS:JW:Entropy:2017}), results with the following matrix:
\begin{gather}
\rho^{T^2}_{A0} = \left(
\begin{array}{cccc}
 \left(\alpha _0\right){}^* \alpha _0 & \imag \left(\sin (t) \beta _0\right){}^* \alpha _0 & \left(\alpha _0\right){}^* \cos (t) \beta _0 & \imag \left(\sin (t) \beta _0\right){}^* \cos (t) \beta _0 \\
 -\imag \left(\alpha _0\right){}^* \sin (t) \beta _0 & \left(\sin (t) \beta _0\right){}^* \sin (t) \beta _0 & 0 & 0 \\
 \left(\cos (t) \beta _0\right){}^* \alpha _0 & 0 & \left(\cos (t) \beta _0\right){}^* \cos (t) \beta _0 & 0 \\
 -\imag \left(\cos (t) \beta _0\right){}^* \sin (t) \beta _0 & 0 & 0 & 0 \\
\end{array}
\right) .
\end{gather}

The eigenvalues are:
\begin{eqnarray}
\lambda_0 & = & -\beta_0 \sqrt{\left(\beta_0\right){}^* \cos\left(t\right)} \sqrt{\sin (t) \cos (t)} \sqrt{\left(\beta_0\right){}^* \sin \left(t\right)}, \notag \\
\lambda_1 & = & \beta_0 \sqrt{\sin (t) \cos (t)} \sqrt{{\left(\beta_0\right)}^* \sin \left(t\right)} \sqrt{\left(\beta_0\right){}^* \cos \left(t\right)}, \notag \\
\lambda_2 & = & \frac{1}{2} \left( 1 - \sqrt{ {\left| \alpha_0 \right|}^4 + 2 {\left| \alpha_0 \beta_0\right|}^2 + {\left| \beta_0 \right|}^4 \cos^2(2t)} \right), \notag \\
\lambda_3 & = & \frac{1}{2} \left( 1 + \sqrt{ {\left| \alpha_0 \right|}^4 + 2 {\left| \alpha_0 \beta_0\right|}^2 + {\left| \beta_0 \right|}^4 \cos^2(2t)} \right).
\end{eqnarray}

\begin{figure}
\begin{center}
\includegraphics[height=3.75cm]{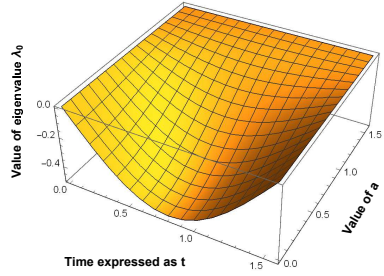}
\end{center}
\caption{The changes of eigenvalue $\lambda_0$ for quantum switch in state $\mket{A01}$, test PPT, time $t \in \langle 0, \frac{\pi}{2} \rangle$}
\label{lbl:fig:L0PPT:changing:EE:MS:JW:EtropyArt2017}
\end{figure}

We obtained the negative eigenvalue $\lambda_0$ -- the changes of this value are presented at Fig.~\ref{lbl:fig:L0PPT:changing:EE:MS:JW:EtropyArt2017}. It should be pointed out that for $t=0$ and $t=\frac{\pi}{2}$ the system of the switch is in a separable state, according to the expectations, because for $t=0$ the switch is in its initial state and for $t=\frac{\pi}{2}$ the switch is in its final state.

The changes of the entanglement's level may be observed with use of Concurrence measure. If we calculate the entanglement's value according to Eq.~(\ref{lbl:eq:concurrency:MS:JW:EntropyArt2017}), we obtain only one non-zero eigenvalue which, after some transformations, allows to formulate the final equation:

\begin{equation}
\mathcal{C}(\mket{A01},t) = \left| \beta_0^2 \sin (2t) \right| .
\label{lbl:eq:concurrency:MS:JW:Entropy:2017}
\end{equation}

\begin{figure}
\begin{center}
\includegraphics[height=3.75cm]{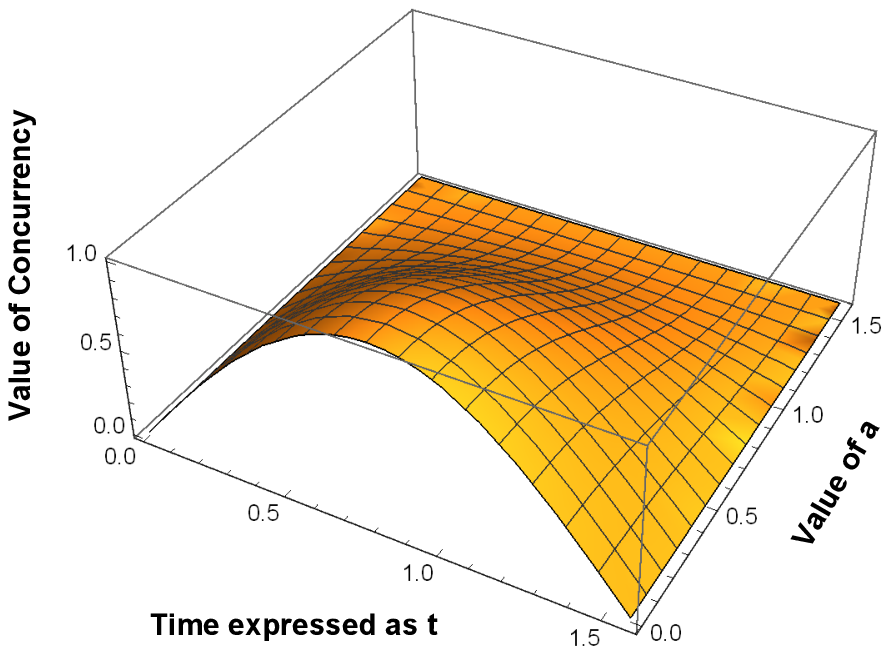}
\end{center}
\caption{The change of Concurrency value for quantum switch in state $\mket{A01}$,  time $t \in \langle 0, \frac{\pi}{2} \rangle$, state $\mket{A}$ is given by Eq.~(\ref{eqn:state:A:MS:JW:Entropy:2017})}
\label{lbl:fig:Concurrence:MS:JW:EtropyArt2017}
\end{figure}

Utilizing I-Concurrency leads to the following relation:
\begin{eqnarray}
\mathcal{IC}(\mket{A01},t) = \sqrt{2} \sqrt{-\left(\left| \alpha_0\right| {}^2+\left| \sin(t) \beta_0\right| {}^2\right){}^2-2 \left| \cos(t) \alpha_0 \beta_0\right| {}^2-\left| \cos (t) \beta_0\right| {}^4+1}.
\label{lbl:eq:iconcurrency:MS:JW:Entropy:2017}
\end{eqnarray}
which depicts the influence of time on the values of probability amplitudes. However, it does not matter if we use approach described by Eq.~(\ref{lbl:eq:concurrency:MS:JW:Entropy:2017}) or (\ref{lbl:eq:iconcurrency:MS:JW:Entropy:2017}).  Fig.~(\ref{lbl:fig:Concurrence:MS:JW:EtropyArt2017}) shows the change of the entanglement's level calculated with use of Concurrency for the state $\mket{A01}$ given by Eq.~(\ref{eqn:state:A:MS:JW:Entropy:2017}).

The value of entropy may be also calculated for the state $\mket{A01}$, then the controlling qubit is removed from the state by the operation of partial transposition. The obtained state $\mket{A0}$ may be presented as a density matrix (\ref{eqn:state:for:Schmidt:decomp:MS:JW:Entropy:2017}). The matrix may be used to determine the value of entropy. Another operation of partial trace performed on the state $\rho_{A0}$ removes the second qubit and then the quantum state may be expressed as the following density matrix:

\begin{equation}
\mptr{A0}{B} =  \left(
\begin{array}{cc}
 \left| \alpha_0 \right| {}^2+\left| \sin(t) \beta_0 \right| {}^2 & \left( \cos(t) \beta_0 \right){}^* \alpha_0 \\
 \left( \alpha_0 \right){}^* (\cos  t) \beta_0 & \left| \cos(t) \beta_0 \right| {}^2 \\
\end{array}
\right) .
\label{lbl:eq:state:for:entropy:MS:JW:Entropy:2017}
\end{equation}

Performing spectral decomposition on the state~(\ref{lbl:eq:state:for:entropy:MS:JW:Entropy:2017}) we will obtain the eigenvalues: 
\begin{eqnarray}
\lambda_0 = \frac{1}{2} \left(1 - \sqrt{2 \left| \alpha_0 \beta _0\right| {}^2+\left| \alpha_0\right| {}^4 + \left| \beta_0 \right| {}^4 \cos ^2(2 t)}\right), \notag \\
\lambda_1 = \frac{1}{2} \left(1 + \sqrt{2 \left| \alpha_0 \beta_0\right| {}^2+\left| \alpha_0 \right| {}^4 + \left| \beta_0 \right| {}^4 \cos^2(2 t)} \right).
\end{eqnarray}

\begin{figure}
\begin{center}
\includegraphics[height=4.75cm]{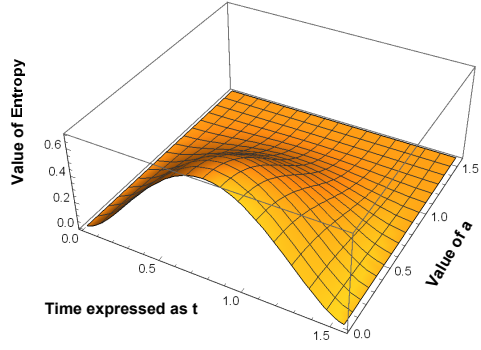}
\end{center}
\caption{The change of entropy value for quantum switch in state $\mket{A01}$, time $t \in \langle 0, \frac{\pi}{2} \rangle$; the state $\mket{A}$ is given by Eq.~(\ref{eqn:state:A:MS:JW:Entropy:2017})}
\label{lbl:fig:Entropy:MS:JW:EtropyArt2017}
\end{figure}

The obtained eigenvalues allow to directly compute the value of entropy. Fig.~(\ref{lbl:fig:Entropy:MS:JW:EtropyArt2017}) shows the changes of entropy values during the work of quantum switch for the state~(\ref{eqn:state:A:MS:JW:Entropy:2017}) in time interval $t \in \langle 0, \frac{\pi}{2} \rangle$.

\section{Quantum switch with noise presence} \label{lbl:qs:noise:presence:MS:JW:Entropy:2017}

Introducing a noise modeled by the quantum channels, described in Table~(\ref{lbl:tbl:quantum:channels:MS:JW:Entropy:2017}), causes the changes in a value of entanglement in an analyzed system. Utilizing I-Concurrency measure we can elaborate formulae expressing the value of entanglement in the moment $t$. We assume that the noise is present on the first qubit of state $\mket{A01}$. For the Phase Flip channel and the state $\mket{A01}$, described by Eq.~(\ref{eqn:state:A:MS:JW:Entropy:2017}), the value of I-Concurrency measure is:
\begin{equation}
\nIC(t,p) = \sqrt{2 - 4(1-2 \text{p})^2 \left| \alpha _0\right|{}^2 \left| \cos(t) \beta _0\right|{}^2-2 \left(\left| \alpha _0\right|{}^2 + \left|\sin (t) \beta_0 \right|{}^2 \right){}^2 - 2 \left| \cos (t) \beta _0\right|{}^4} .
\label{lbl:eq:IC:PF:MS:JW:EtropyArt2017}
\end{equation}

In the case of Bit Flip channel we obtain:
\begin{multline}
\nIC(t,p) = \left[2 - 2\left( p \left| \cos (t) \beta _0 \right|{}^2-( p - 1 ) \left( \left| \alpha_0 \right|{}^2 + \left| \sin (t) \beta_0 \right| {}^2\right)\right){}^2 -  \right. \\
2 \left. \left( (p-1) \left| \cos (t) \beta _0 \right|{}^2-p \left( \left| \alpha_0 \right|{}^2 + \left| \sin (t) \beta_0 \right| {}^2\right)\right){}^2 - \right. \\
\left. 4 \left(\alpha_0 p \left(\cos(t) \beta_0\right){}^* - \beta_0 (p-1) \left(\alpha_0\right){}^* \cos(t)\right) \right. \\
\left. \left(\beta_0 p \left(\alpha_0\right){}^* \cos(t) - \alpha_0 (p-1) \left(\cos(t) \beta_0\right){}^*\right) \right]^{1/2} .
\label{lbl:eq:IC:BF:MS:JW:EtropyArt2017}
\end{multline}

The value of entanglement when Amplitude Damping channel was used is:
\begin{multline}
\nIC(t,p) = \left[2 + 4 (p-1) \left| \alpha_0\right| {}^2 \left| \cos (t) \beta_0\right| {}^2-2 \left(\left| \alpha_0\right| {}^2 + p \left| \cos (t) \beta_0 \right| {}^2+\left| \sin (t) \beta_0\right| {}^2\right){}^2 - \right. \\
\left. 2 (p-1)^2 \left| \cos (t) \beta_0 \right|{}^4\right]^{1/2} .
\label{lbl:eq:IC:AD:MS:JW:EtropyArt2017}
\end{multline}

And finally for Phase Damping channel the value of I-Concurrency measure is:
\begin{equation}
\nIC(t,p) = \sqrt{2 + 4 (p-1) \left| \alpha_0 \right|{}^2 \left| \cos (t) \beta_0 \right|{}^2-2 \left(\left| \alpha_0 \right|{}^2+\left| \sin (t) \beta_0 \right| {}^2\right){}^2-2 \left| \cos (t) \beta_0\right|{}^4} .
\label{lbl:eq:IC:PD:MS:JW:EtropyArt2017}
\end{equation}

\begin{figure}
\begin{center}
\begin{tabular}{|c|c|}
\hline
Conccureny with Noise & Difference \\ \hline
\multicolumn{2}{|c|}{Phase Flip} \\  \hline
& \\
\includegraphics[height=3.75cm]{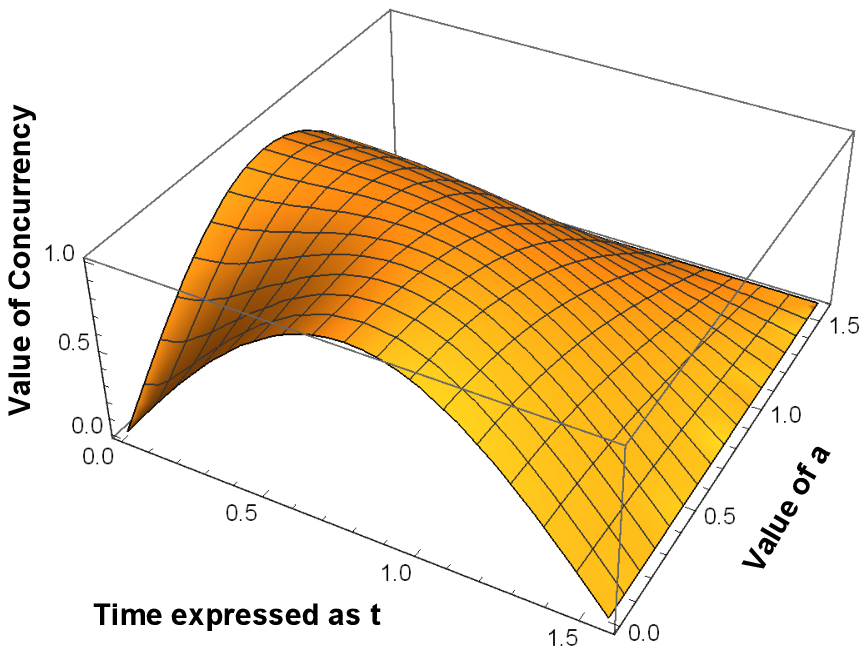} & \includegraphics[height=3.75cm]{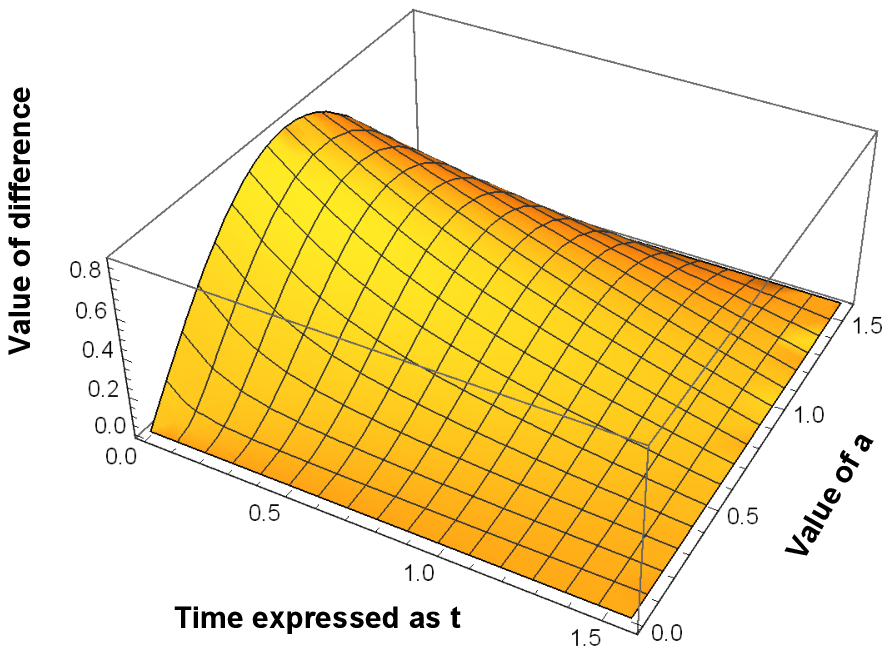} 
\\
& \\ \hline
\multicolumn{2}{|c|}{Bit Flip} \\  \hline
& \\  \hline
\includegraphics[height=3.75cm]{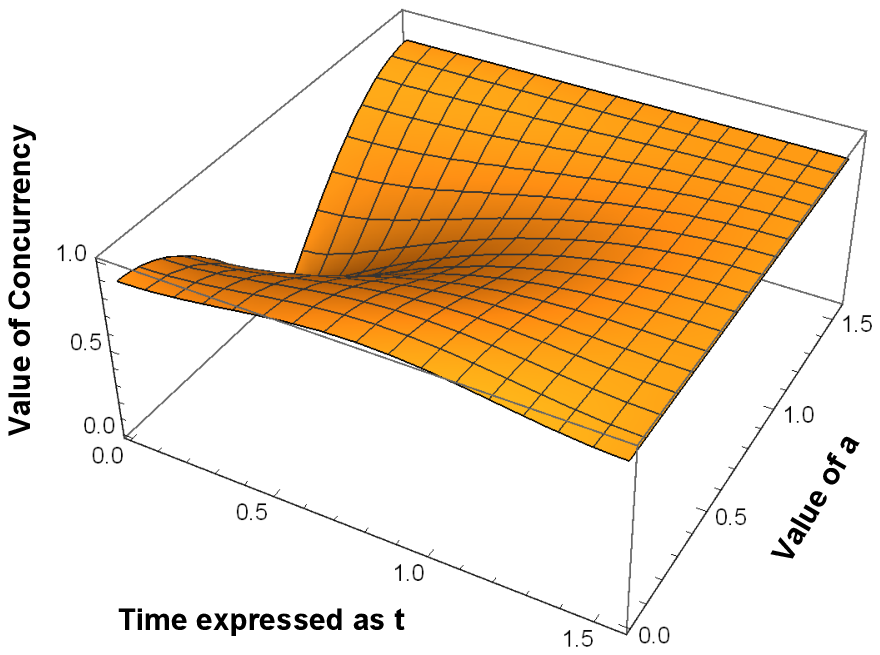} & \includegraphics[height=3.75cm]{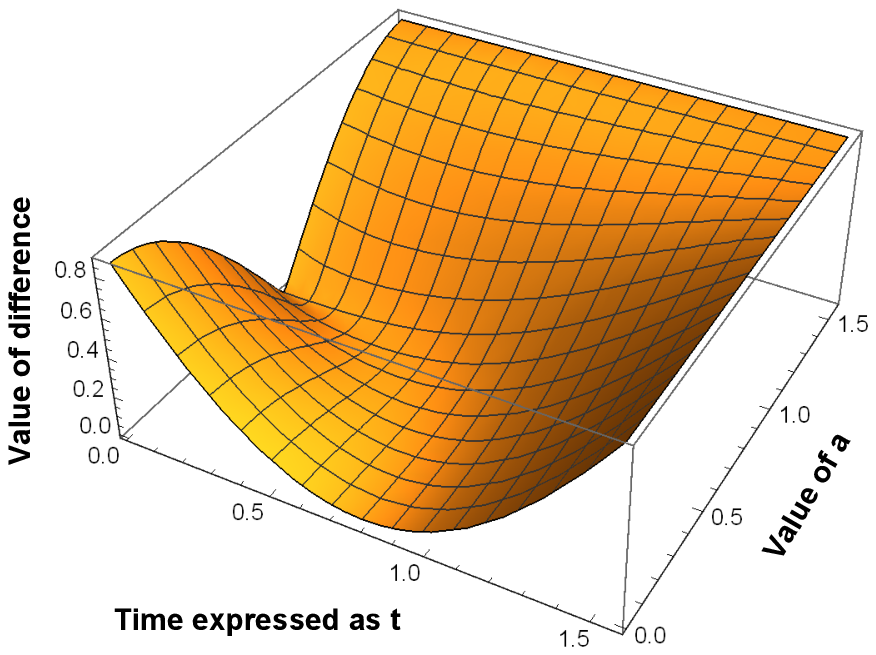} 
\\ \hline
& \\ \hline
\multicolumn{2}{|c|}{Amplitude Damping} \\  \hline
& \\  \hline
\includegraphics[height=3.75cm]{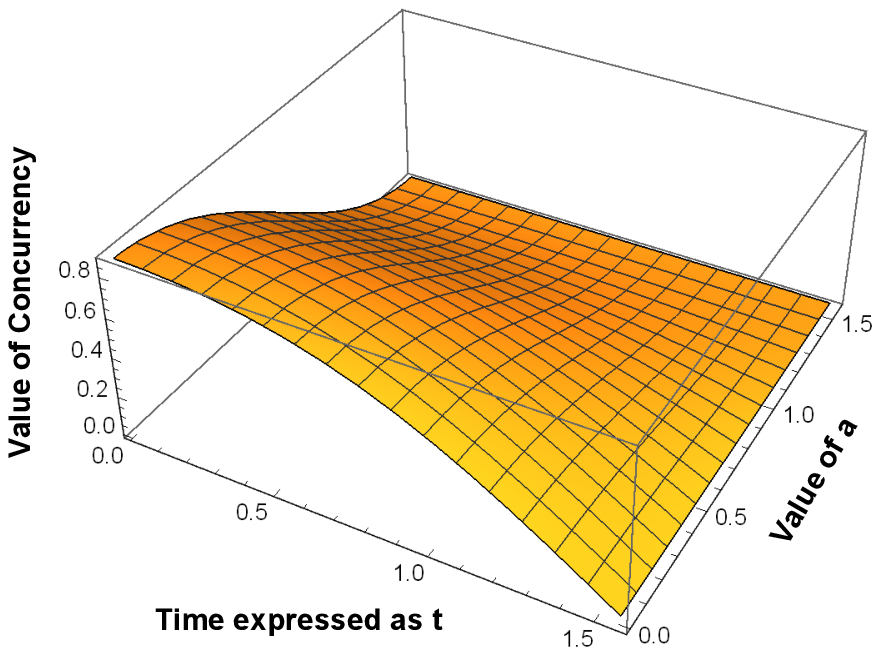} & \includegraphics[height=3.75cm]{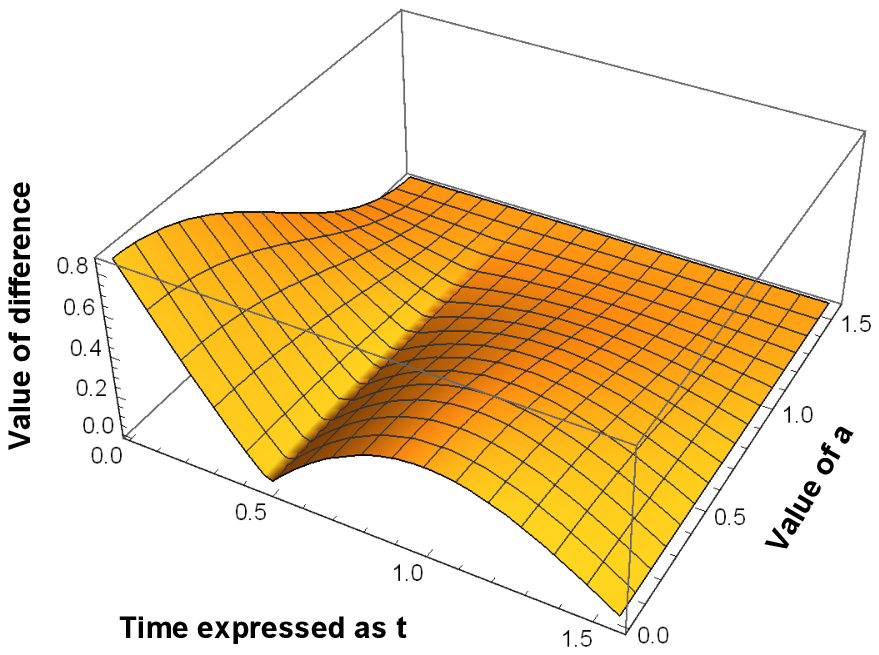} 
\\ \hline
& \\  \hline
\multicolumn{2}{|c|}{Phase Damping} \\  \hline
& \\  \hline
\includegraphics[height=3.75cm]{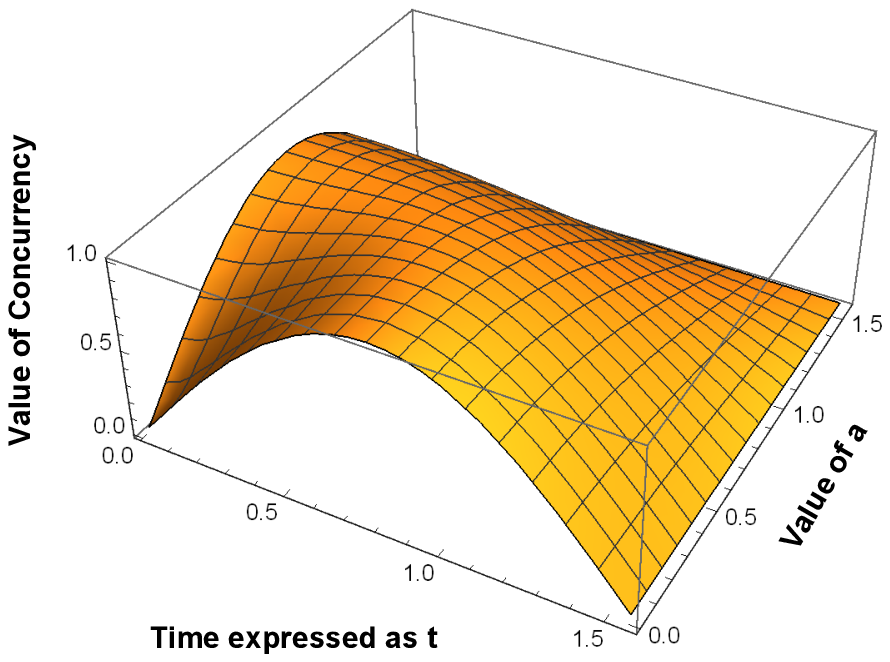} & \includegraphics[height=3.75cm]{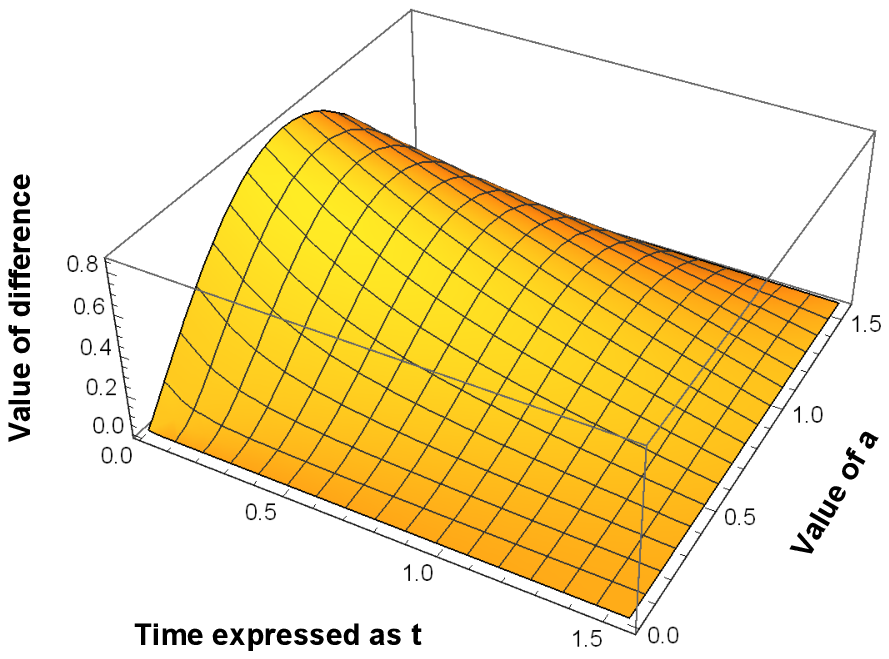}
\\ \hline
\end{tabular}
\end{center}
\caption{The changes of I-Concurrency value for the quantum switch working on state $\mket{A01}$ with four types of noise. The time interval is $t \in \langle 0, \frac{\pi}{2} \rangle$ and the state $\mket{A}$ is given by Eq.~(\ref{eqn:state:A:MS:JW:Entropy:2017}). Probability of decoherence $p=0.74$. The charts in the column Difference show the absolute difference between the values of I-Concurrency measure for the systems without noise and with noise on the first qubit
}
\label{lbl:fig:Concurrence:and:Noise:MS:JW:EtropyArt2017}
\end{figure}

Utilizing the relations (\ref{lbl:eq:IC:PF:MS:JW:EtropyArt2017}), (\ref{lbl:eq:IC:BF:MS:JW:EtropyArt2017}), (\ref{lbl:eq:IC:AD:MS:JW:EtropyArt2017}), (\ref{lbl:eq:IC:PD:MS:JW:EtropyArt2017}) we can obtain a visualization of changes in the level of entanglement e.g. on the first qubit. In each case the level of entanglement was changed. The noise was added only to the first qubit so the observed changes appear in the first phase of switch's work. Fig.~(\ref{lbl:fig:Concurrence:and:Noise:MS:JW:EtropyArt2017}) depicts the changes of entanglement's value and also the difference in comparison with the work of switch on a pure state (i.e. without any noise).

\begin{remark}
It should be emphasized that according to (\ref{lbl:eq:IC:PF:MS:JW:EtropyArt2017}) and (\ref{lbl:eq:IC:PD:MS:JW:EtropyArt2017}) there are some differences between the values of entanglement calculated with use of Concurrency measure, even if the charts at Fig.~(\ref{lbl:fig:Concurrence:and:Noise:MS:JW:EtropyArt2017}) do not depict these differences.
\end{remark}

\begin{figure}
\begin{center}
\begin{tabular}{|c|c|}
\hline
Entropy with Noise & Difference \\ \hline
\multicolumn{2}{|c|}{Phase Flip} \\  \hline
& \\
\includegraphics[height=3.75cm]{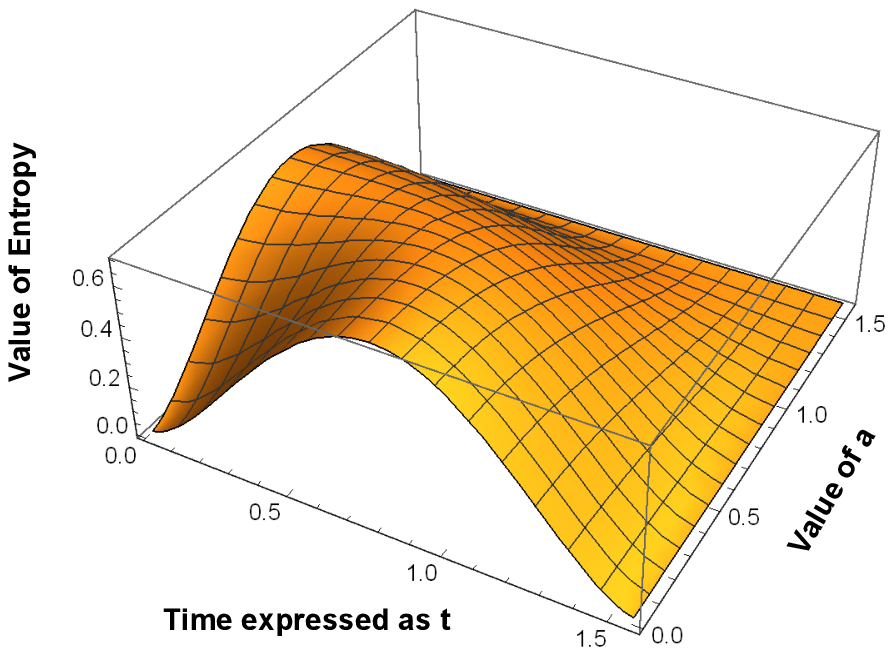} & \includegraphics[height=3.75cm]{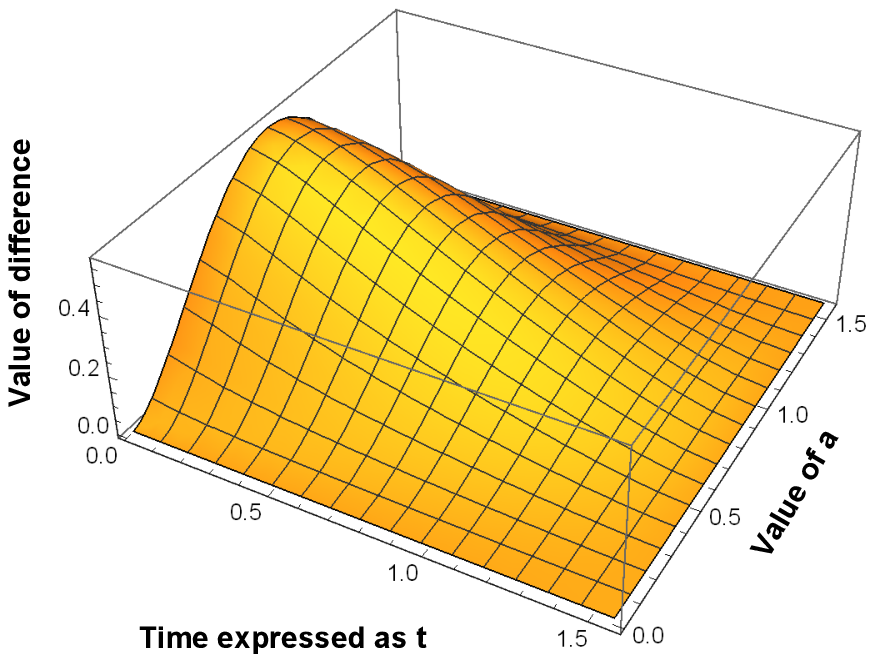} 
\\
& \\ \hline
\multicolumn{2}{|c|}{Bit Flip} \\  \hline
& \\  \hline
\includegraphics[height=3.75cm]{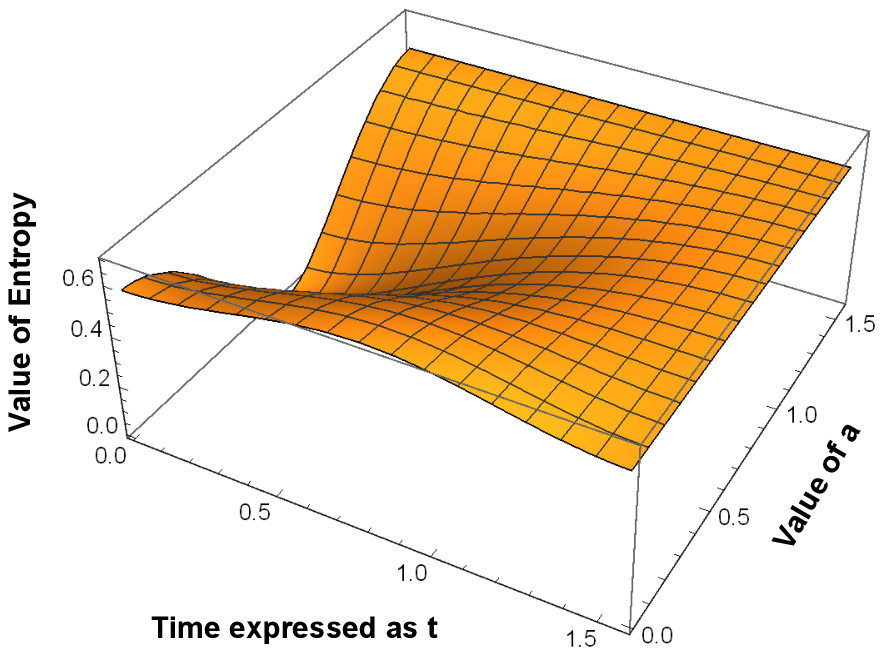} & \includegraphics[height=3.75cm]{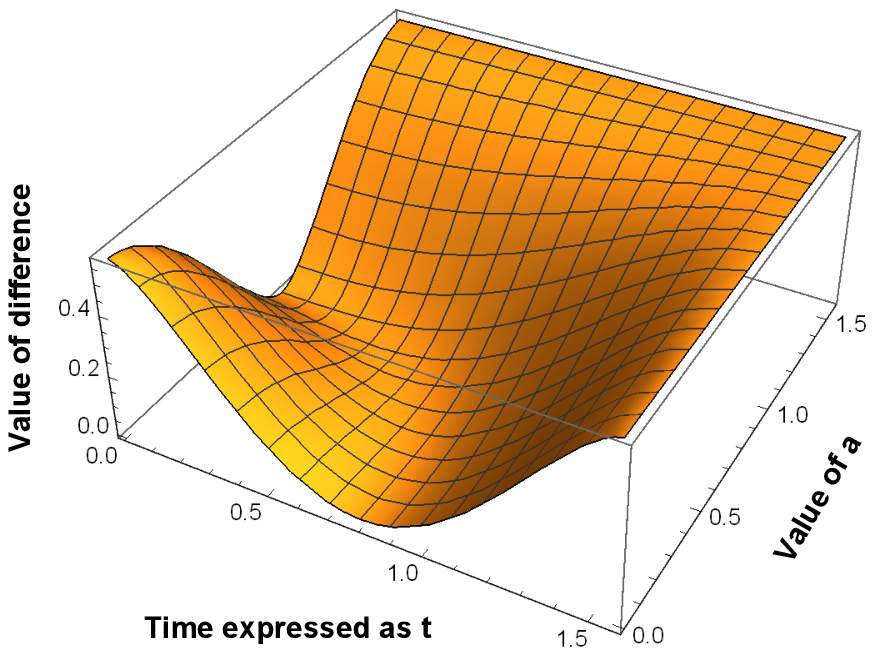} 
\\ \hline
& \\ \hline
\multicolumn{2}{|c|}{Amplitude Damping} \\  \hline
& \\  \hline
\includegraphics[height=3.75cm]{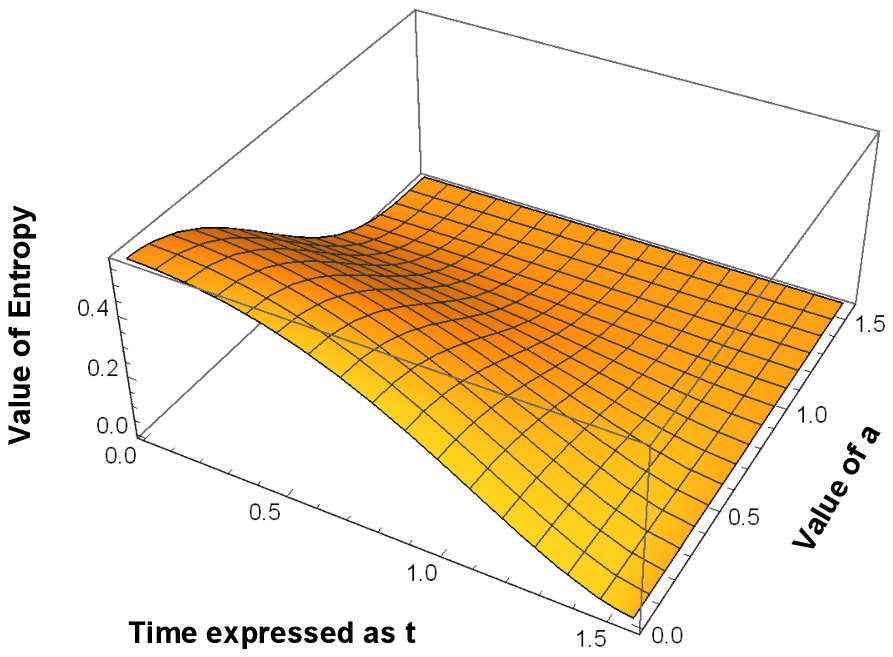} & \includegraphics[height=3.75cm]{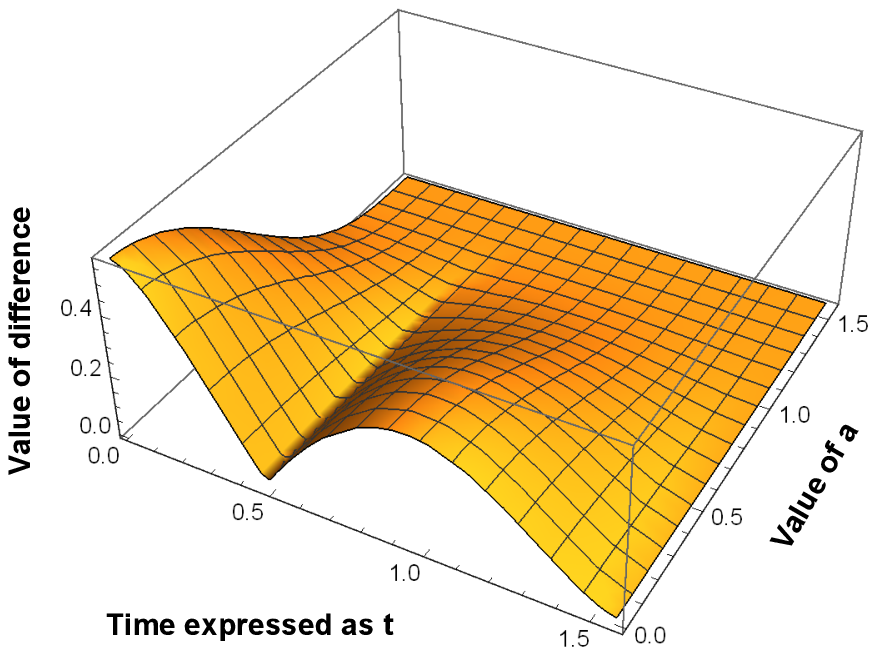} 
\\ \hline
& \\  \hline
\multicolumn{2}{|c|}{Phase Damping} \\  \hline
& \\  \hline
\includegraphics[height=3.75cm]{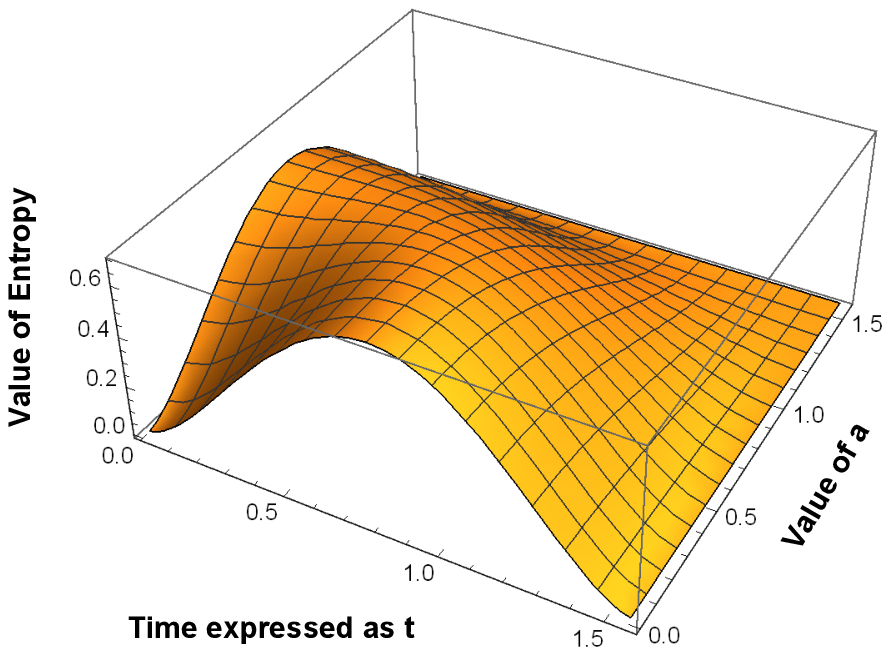} & \includegraphics[height=3.75cm]{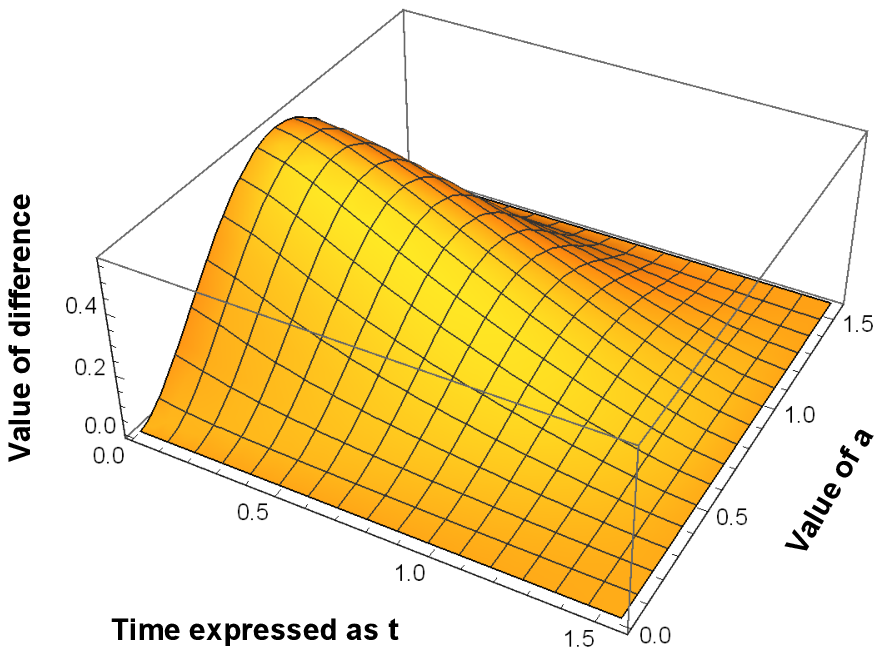} 
\\ \hline
\end{tabular}
\end{center}
\caption{The changes of Entropy value for the quantum switch working on state $\mket{A01}$ with four types of noise. The time interval is $t \in \langle 0, \frac{\pi}{2} \rangle$ and the state $\mket{A}$ is given by Eq.~(\ref{eqn:state:A:MS:JW:Entropy:2017}). Probability of decoherence $p=0.74$. The charts in the column Difference show the absolute difference between the values of Entropy for the systems without noise and with noise on the first qubit
}
\label{lbl:fig:Entropy:and:Noise:MS:JW:EtropyArt2017}
\end{figure}

The values of entropy also correctly point the levels of entanglement as it is shown at Fig.~(\ref{lbl:fig:Entropy:and:Noise:MS:JW:EtropyArt2017}).	The obtained values, though different in the numerical aspect, allow to detect if the switch works properly or the noise is present.  

It should be emphasized that the charts placed at Fig.~(\ref{lbl:fig:Concurrence:and:Noise:MS:JW:EtropyArt2017}) and  Fig.~(\ref{lbl:fig:Entropy:and:Noise:MS:JW:EtropyArt2017}) depict the noise affecting the first qubit of analyzed quantum state. That explains why the changes in the value of entanglement's level are present in the first working phase of the system. It is especially visible for the Amplitude Damping channel where the entanglement is present from the very beginning of the computational process. The same phenomenon may be seen when the noise affects the second qubit -- then the changes of entanglement's level occur in the second working phase of the switch.

\subsection{Average Fidelity}
The evaluation of noise influence on the work of quantum switch may be also calculated as an average value of Fidelity measure.
The works \cite{Nielsen2002}, \cite{Pedersen2007}, \cite{Bowdrey2002} show that the average value of Fidelity may be computed with the following equation:
\begin{equation}
F_{avg}(U_0, \mathcal{E}) = \frac{1}{n(n+1)} \left( \mtr{\sum_k M^{\dagger}_{k}M_k} + \sum_k {| \mtr{M_k} |}^2 \right),
\end{equation}
where $M_k = {U_t}^{\dagger} E_k$ and $\mathcal{E}(\rho) = \sum_{k} E_k \rho E^{\dagger}_k$ represent utilized type of noise as a quantum channel.

The unitary operation $U_{qs}(t)$ presented in Eq.~(\ref{lbl:eqn:u:qs:oper}) may be used to examine the influence of each quantum channel on the switch's efficiency -- the appropriate formulae are gathered in Table~(\ref{lbl:tbl:average:fidelity:MS:JW:Entropy:2017}). 

\begin{table}
\begin{displaymath}
\begin{array}{|l|c|}
\hline
\multicolumn{1}{|c|}{\mathrm{Name \; of \; channel}} & Formula  \\  \hline
\mathrm{Phase \; Flip (PF)}  & F_{avg} (p,t) = \frac{1}{18} \left(p (\cos (t)+3)^2+2\right) \\ \hline
\mathrm{Bit \; Flip (BF)}  & F_{avg} (p,t) = \frac{1}{18} \left( p  (\cos (t)+3)^2 + 2 \right) \\ \hline
\mathrm{Amplitude \; Damping (AD)}  & F_{avg} (p,t) = \frac{1}{72} \left( \left| \left(\sqrt{1-p}+1\right) \left(\cos(t)+3\right) \right|^2 + 8 \right) \\ \hline
\mathrm{Phase \; Damping (PD)}  &  F_{avg} (p,t) = \frac{1}{72} \left(\left| \left(\sqrt{1-p}+1\right) (\cos (t)+3)\right| ^2+\left| p (\cos (t)+3)^2\right| +8\right) \\ \hline
\end{array}
\end{displaymath}
\caption{The formulae describing the average value of Fidelity respectively to the probability of noise presence $p \in \langle 0, 1 \rangle$ and time $t \in \langle 0, \frac{\pi}{2}\rangle$ (the average value of Fidelity is equal for PF and BF channels). The noise applies to the first qubit of state $\mket{A01}$
}
\label{lbl:tbl:average:fidelity:MS:JW:Entropy:2017}
\end{table}

\begin{figure}
\begin{tabular}{|c|c|}
\hline
Phase Flip & Bit Flip \\ \hline 
& \\ 
\includegraphics[height=3.75cm]{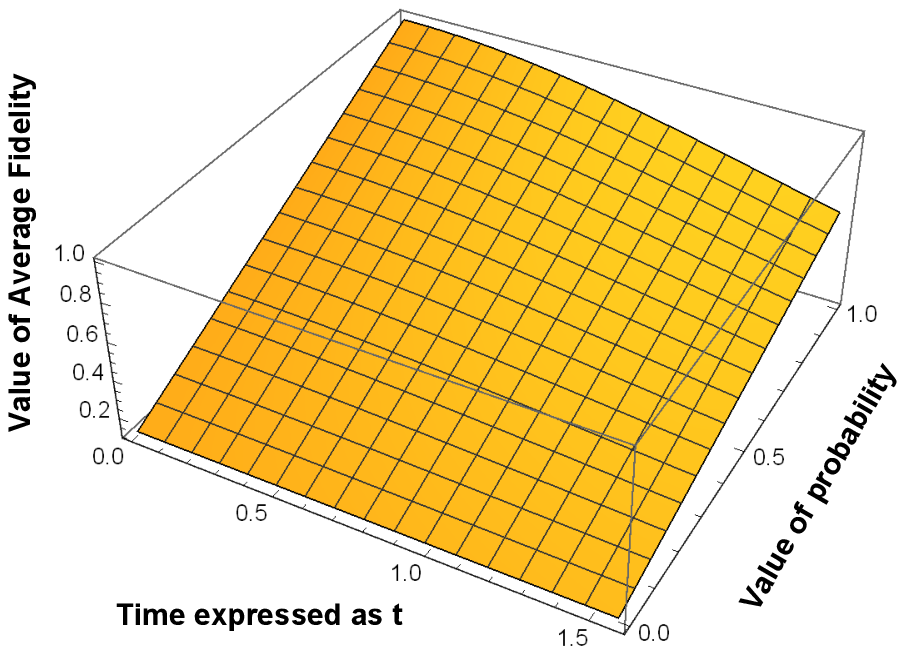} & \includegraphics[height=3.75cm]{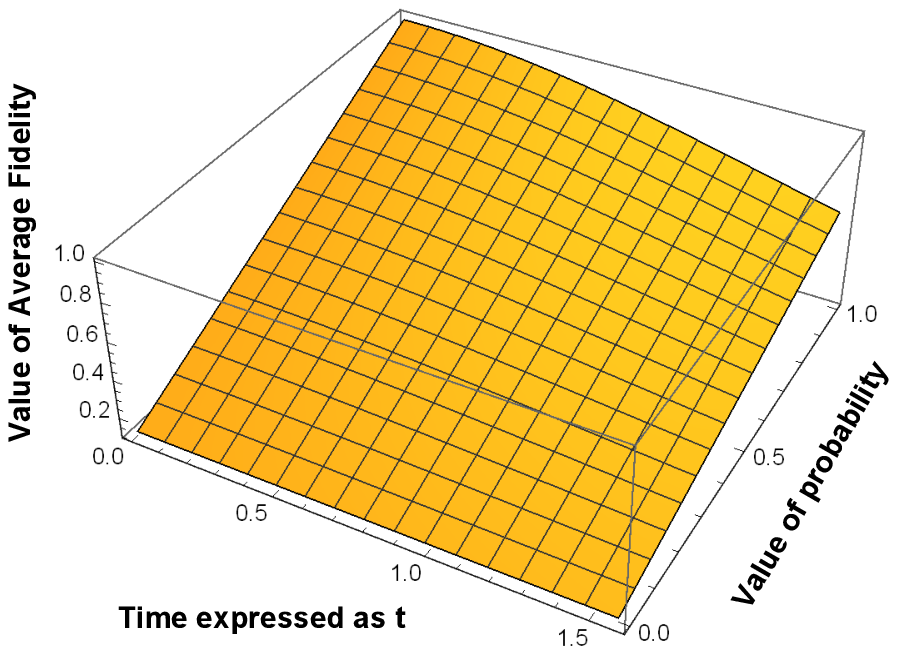} 
\\ \hline
Amplitude Damping & Phase Damping \\
& \\  \hline
\includegraphics[height=3.75cm]{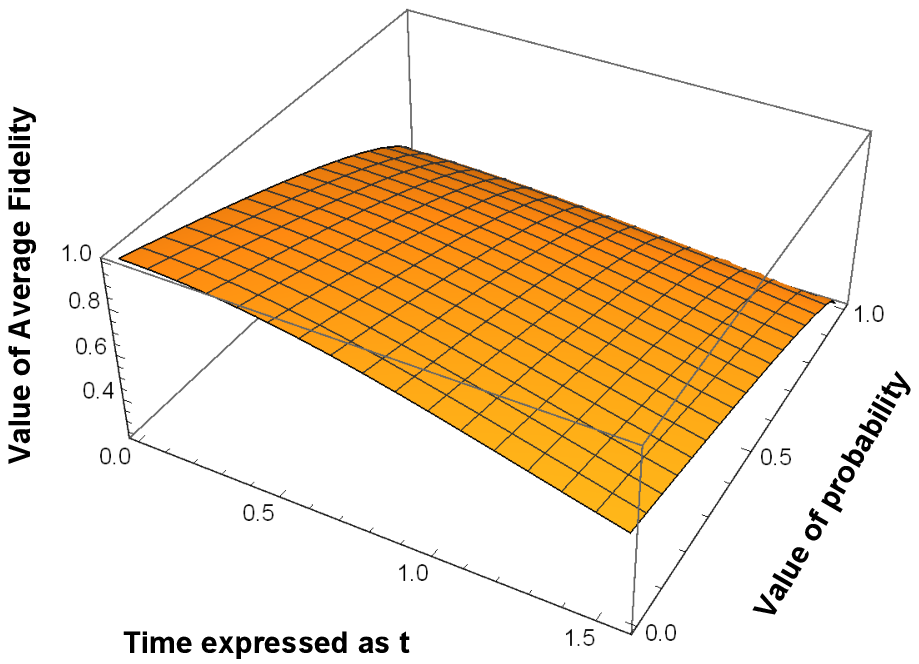} & \includegraphics[height=3.75cm]{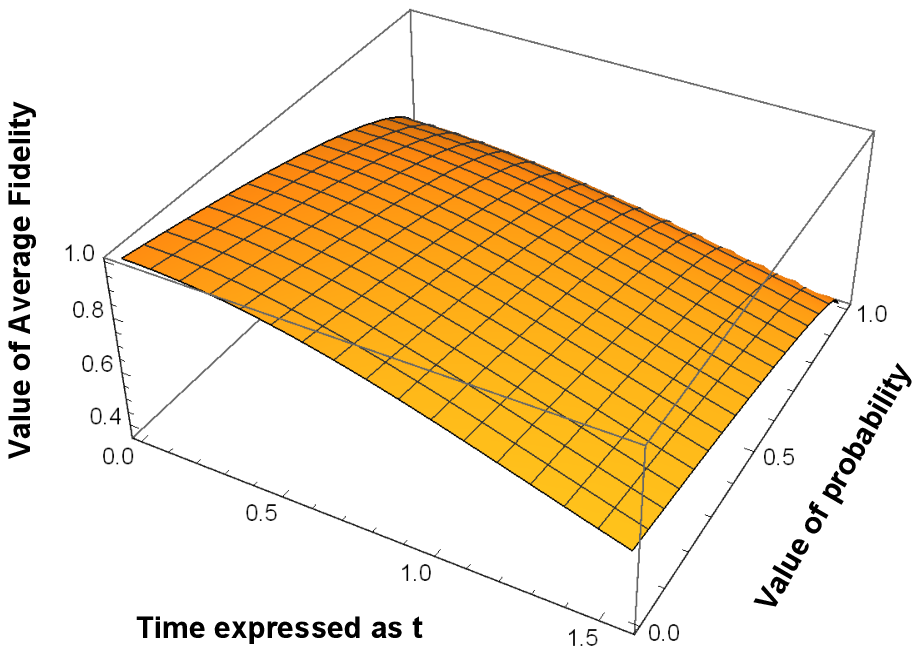}  \\ \hline
\end{tabular}
\caption{The average value of Fidelity measure for the state $\mket{A01}$ with four different noise channels, time $t \in \langle 0, \frac{\pi}{2} \rangle$, the state $\mket{A}$ is given by Eq.~(\ref{eqn:state:A:MS:JW:Entropy:2017})}
\label{lbl:fig:Avg:Fid:MS:JW:EtropyArt2017}
\end{figure}

According to our expectations, the noise introduced by the channels described in Table~\ref{lbl:tbl:quantum:channels:MS:JW:Entropy:2017} decreases the average value of Fidelity measure during the switch's work. This reduction is quite significant because the quantum state processed by the switch is seriously disordered when the probability of noise presence $p > 0.1$. Fig.~(\ref{lbl:fig:Avg:Fid:MS:JW:EtropyArt2017}) shows the average values of Fidelity for the time variable $t$. The charts clearly depict the regular decreasing of average value of Fidelity in every quantum channel. 

\section{Conclusions} \label{lbl:conclusions:MS:JW:Entropy:2017}

In this paper we analyzed the entanglement's level during the work of quantum switch. Duration of the experiment is represented by variable $t$ which accepts values from range of real numbers: $\langle 0, \frac{\pi}{2} \rangle$.

An analysis of the entanglement level with use of Schmidt decomposition and PPT criterion confirms the presence of entanglement during the work of quantum switch. If the switch operates correctly, i.e. without any noise, there is no entanglement at the beginning and at the end of the process. Introducing noise to the system changes the level of entanglement. That allows to evaluate if in the moment $t$ the switch works properly.  

We assume that approach presented in this work allows to detect incorrect operating the switch when the system is under an influence of noise. Evaluating the level of entanglement also allows to verify if the quantum circuit realizing the switch is correct. Taking into account this aspect, we can state that the level of entanglement provides an information about correctness of used algorithm. Further work concerning this issue may help to develop the research area reffering to the correctness of the quantum algorithms \cite{Ying2013}, i.e.  loop quantum theory \cite{Ying2010}, weakest quantum precondition \cite{Gielerak2010}.

\section*{Acknowledgments}

{\small We would like to thank for useful discussions with the~\textit{Q-INFO} group at the Institute of Control and Computation Engineering (ISSI) of the University of Zielona G\'ora, Poland. We would like also to thank to anonymous referees for useful comments on the preliminary version of this paper. The numerical results were done using the hardware and software available at the ''GPU/FPGA $\mu$-Lab'' located at the Institute of Control and Computation Engineering of the University of Zielona G\'ora, Poland. }

\end{document}